\documentclass[12pt]{article}

\usepackage{cancel,slashed}
\usepackage{bbm}
\usepackage{amssymb}
\usepackage{amsfonts}
\usepackage{amsmath}
\usepackage{graphicx}
\usepackage{cite}
\usepackage{latexsym}
\usepackage{color}
\usepackage{xypic}
\usepackage{cancel,slashed}
\usepackage[linktocpage]{hyperref}
\usepackage{color}
\usepackage{transparent}

\newcommand{\bmat}{\left(\begin{array}}
\newcommand{\emat}{\end{array}\right)}

\def\p{\partial}
\def\a{\alpha}

\def\b{\beta}

\def\d{\delta}
\def\th{\theta}

\def\Om{\Omega}

\def\-{\hphantom{-}}

\def\s2{\frac{1}{\sqrt2}}

\def\oh{\frac{1}{2}}
\def\beq{\begin{equation}}
\def\eeq{\end{equation}}
\def\beqa{\begin{eqnarray}}
\def\eeqa{\end{eqnarray}}

\def\tr{{\rm tr \,}}

\def\T{{\rm T}}
\def\Z{{\mathbb Z}}

\def\cw{{\mathcal W}} 

\def\cn{{\mathcal N}}
\def\ct{{\mathcal T}}

\def\Dsl{\,\raise.15ex\hbox{/}\mkern-13.5mu D} 

\def\CM {{\cal M}}
\def\CR {{\cal R}}
\def\CN {{\cal N}}

\def\CL {{\cal L}}

\def\tr{\mbox{Tr}}

\def\be{\begin{equation}}
\def\ee{\end{equation}}
\def\bea{\begin{eqnarray}}
\def\eea{\end{eqnarray}}
\def\raw{\rightarrow}

\def\IZ{\mathbb{Z}}

\def\Id{{\mathbb{I}}}

\def\T{{\bf T}}

\def\oh{\frac{1}{2}}
\def\a{{\alpha}}
\def\b{{\beta}}
\def\d{{\delta}}
\def\eps{{\epsilon}}
\def\th{{\theta}}

\def\lam{{\lambda}}
\def\Om{{\Omega}}

\def\G{{\Gamma}}

\def\p{{\partial}}

\def\vec#1{{\overrightarrow{#1}}}

\def\sm2{{\mbox{\small 2}}}

\newcommand{\drawsquare}[2]{\hbox{%
\rule{#2pt}{#1pt}\hskip-#2pt
\rule{#1pt}{#2pt}\hskip-#1pt
\rule[#1pt]{#1pt}{#2pt}}\rule[#1pt]{#2pt}{#2pt}\hskip-#2pt
\rule{#2pt}{#1pt}}
\newcommand{\Ysymm}{\raisebox{-.5pt}{\drawsquare{6.5}{0.4}}\hskip-0.4pt%
         \raisebox{-.5pt}{\drawsquare{6.5}{0.4}}}
\newcommand{\Yasymm}{\raisebox{-3.5pt}{\drawsquare{6.5}{0.4}}\hskip-6.9pt%
        \raisebox{3pt}{\drawsquare{6.5}{0.4}}}

\begin{document}
\pagestyle{plain}

\makeatletter
\@addtoreset{equation}{section}
\makeatother
\renewcommand{\theequation}{\thesection.\arabic{equation}}
\pagestyle{empty}
\rightline{ IFT-UAM/CSIC-13-065}
\vspace{0.5cm}
\begin{center}
\LARGE{{Discrete Flavor Symmetries in D-brane models}
\\[10mm]}
\large{Fernando Marchesano,$^1$ Diego Regalado$^{1,2}$ and Liliana V\'azquez-Mercado$^3$ \\[10mm]}
\small{
${}^1$ Instituto de F\'{\i}sica Te\'orica UAM/CSIC, Cantoblanco, 28049 Madrid, Spain \\
${}^2$ Departamento de F\'{\i}sica Te\'orica, 
Universidad Aut\'onoma de Madrid, 
28049 Madrid, Spain \\
${}^3$ Departamento de F\'isica, DCI, Campus Le\'on, Universidad de Guanajuato,\\[-0.3em] 
C.P.  37150 Guanajuato, M\'exico.
\\[8mm]} 
\small{\bf Abstract} \\[5mm]
\end{center}
\begin{center}
\begin{minipage}[h]{15.0cm} 

We study the presence of discrete flavor symmetries in D-brane models of particle physics.
By analyzing the compact extra dimensions of these models one can determine when such 
symmetries exist both in the context of intersecting and magnetized D-brane constructions. 
Our approach allows to distinguish between approximate and exact discrete symmetries, and 
it can be applied to compactification manifolds with continuous isometries 
or to manifolds that only contain discrete isometries, like Calabi-Yau three-folds. 
We analyze in detail the class of rigid D-branes models based on a $\IZ_2 \times \IZ_2'$ 
toroidal orientifold, for which the flavor symmetry group is either the dihedral group 
$D_4$ or tensor products of it.  We construct explicit Pati-Salam examples in which 
 families transform in non-Abelian representations of the flavor symmetry group, 
constraining Yukawa couplings beyond the effect of massive U(1) D-brane symmetries.

\end{minipage}
\end{center}
\newpage
\setcounter{page}{1}
\pagestyle{plain}
\renewcommand{\thefootnote}{\arabic{footnote}}
\setcounter{footnote}{0}


\tableofcontents


\section{Introduction}

Discrete flavor symmetries are often invoked in the particle physics literature in order to explain different patterns of quark and lepton masses and mixings, as well as to constrain the flavor structure of supersymmetric extensions of the Standard Model. In a purely field theory approach, one may simply consider the whole set of possible discrete family-dependent symmetries compatible with the Standard Model or its 4d field theory extensions, and then analyze in detail those leading to interesting physics \cite{revdfs}. One may then try to obtain a more geometric understanding of such symmetries via higher-dimensional field theories (see e.g. \cite{afl06}), identifying the flavor symmetries with the discrete symmetries of the compactified extra dimensions. 

Alternatively, one may look for the presence of discrete flavor symmetries in particle physics models based on string theory, aiming for a microscopic description of their origin. For the case of heterotic orbifolds this has been addressed in \cite{heterorbi}, obtaining a classification of possible flavor symmetries in this particular context. In general, one expects that the more restrictive framework of string theory will select a limited number of flavor symmetry groups that are compatible with a realistic particle physics model, as well as specific representations for its matter fields. 

Exploring this more restricted scenario is however not the only motivation to realize discrete flavor symmetries in string theory. Given a 4d particle physics model embedded into string theory, one should be able to determine if a flavor symmetry is exact or approximate. If a symmetry is  exact, and because string theory includes quantum gravity, there are strong arguments indicating that it must be realized as a 4d gauge symmetry (see e.g. \cite{klls95,bs10} and references therein). If on the contrary a symmetry is only approximate, string theory should provide a well-defined answer for the scale at which this symmetry is broken, by which mechanism, and how does the breaking affect the couplings of the 4d effective theory. 

The study of discrete gauge symmetries in string theory has been recently undertaken in a series of papers \cite{cim11,bisu11,bcmru12,bcmu12,bmru13}, in which the basic strategy has been to embed the discrete gauge symmetry in a continuous one that is typically broken at the string scale.\footnote{See \cite{Ibanez:2012wg,Anastasopoulos:2012zu,Honecker:2013hda} for applications to specific string theory models.} From all the stringy setups that have been analyzed in these references a particularly interesting one in terms of flavor is given by models of magnetized D-branes in toroidal compactifications. There the above strategy allows to derive a 4d effective Lagrangian with a manifest discrete flavor gauge symmetry \cite{bcmru12}, and the continuous group in which the discrete symmetry is embedded involves the continuos isometries of the toroidal background. 

The purpose of this paper is to study from a more general perspective the presence of discrete flavor symmetries in D-brane models, using a more direct criterion that does not involve embedding the discrete flavor symmetry in any continuous group. The reason to do so is that generic D-brane vacua involve compactification manifolds without any continuous isometry (like Calabi-Yau threefolds) and there the analysis made in \cite{bcmru12} does not apply. Now, while continuous isometries are absent in them, Calabi-Yau manifolds do possess discrete isometries, which will translate into discrete gauge symmetries of the 4d effective action. In addition, in manifolds with non-trivial torsional one-cycles one will obtain 4d discrete gauge symmetries arising from discrete transformations of the NSNS B-field.\footnote{Moreover, in type II string theory models discrete gauge symmetries are obtained from reducing a RR $(p+1)$-form on a Calabi-Yau manifold $\CM$ with torsional $p$-cycles, that is such that ${\rm Tor\, }H_p(\CM, \IZ)\neq 0$ \cite{cim11}. In the following we will ignore these RR discrete gauge symmetries, since typically only very massive states unrelated to the Standard Model fields are charged under them.} As we will now argue, the flavor symmetry group of a D-brane model can be understood in terms of a subgroup of this Calabi-Yau discrete gauge symmetry group. 

When building a D-brane model one considers D$(p+3)$-branes filling up 4d space-time and wrapping specific $p$-cycles of the Calabi-Yau manifold $\CM$. Then, while the action of the isometry group leaves $\CM$ invariant, it may not leave invariant the set of $p$-cycles that the D-branes wrap. In that case the Calabi-Yau gauge symmetry group generated by isometries will be broken down to the subgroup that also leaves invariant the D-brane content of the model. A similar statement can be made for the gauge symmetries generated by B-field transformations. In general, given a D-brane model in a Calabi-Yau, a 4d discrete symmetry group will arise from those transformations of the metric and B-field that leave invariant {\em both} the closed and open string backgrounds of the compactification. The open string zero modes stretched between the background D-branes will transform non-trivially under transformations of the B-field and metric, and so this discrete gauge symmetry will act as a flavor symmetry group in the 4d effective theory. Finally, from the results of \cite{bcmru12} one can see that in the presence of D-branes the action of the generators of discrete isometries and B-field transformations do not commute, and so one typically ends up with a non-Abelian discrete flavor symmetry group generated by these bulk transformations.

While it is non-trivial to construct explicit Calabi-Yau models, one can illustrate all the above statements by means of  D-brane models on toroidal orbifold backgrounds. Toroidal orbifolds are very simple examples of compactification manifolds with discrete isometries, which are realised as certain permutations of the orbifold fixed points. Still, they have been shown to be a fruitful framework to construct semi-realistic D-brane models of particle physics, and specially by considering models of intersecting D6-branes \cite{thebook,reviews}. 

For concreteness we will focus our discussion in a particular orbifold background, namely the type IIA $\IZ_2 \times \IZ_2'$ orientifold with rigid intersecting D6-branes analyzed in \cite{bcms05}. While quite simple, this class of models allows to construct several semi-realistic examples with non-trivial flavor symmetries based on the dihedral group $D_4$, as we show explicitly. We also find that, given a set of D6-branes it is straightforward to detect the presence of a discrete flavor symmetry for them in terms of their pairwise intersection numbers. This simple description allows in turn to characterize a useful notion of exact and approximate discrete symmetry, as we briefly discuss and illustrate via explicit examples. 

An interesting feature of this background is that it has a simple dual description in terms of a type I orbifold with internal magnetic fluxes \cite{aadds99,dt05}, to which our approach to detect flavor symmetries can also be applied. This T-dual description allows to make direct contact with the results of \cite{ako08,acko08,acko09,acko09b,acko10}. In \cite{acko09} the presence of discrete flavor symmetries were detected by analyzing the zero mode wavefunctions of magnetized D-brane models \cite{magnus}. In our approach wavefunctions are not necessary to detect the flavor symmetry, but they are still important to compute the transformation properties of matter fields. We perform a general analysis of the possible family representations valid for both intersecting and magnetized D-brane models, obtaining agreement with previous results. 

This paper is organized as follows. In section \ref{s:review} we review the basic features of the D6-brane models of \cite{bcms05}, and reproduce the chiral index between two D6-branes with a different approach. In section \ref{s:intersec} we discuss the appearance of discrete flavor symmetries in this background using a simple geometric description, which allows to characterize the notion of approximate flavor symmetry. In section \ref{s:magnet} we turn to the dual framework of magnetized D-brane models, reproducing the same flavor symmetry group and classifying the different family representations. We illustrate all the above results via a couple of Pati-Salam examples in section \ref{s:example}, and leave our conclusions for section \ref{s:conclusions}.

Several technical details have been relegated to the appendices. Appendix \ref{ap:review} contains more technical results of  \cite{bcms05} on the $\IZ_2 \times \IZ_2'$ orientifold background. Appendix \ref{ap:inter} derives the effective Lagrangian describing the discrete flavor symmetries for intersecting and magnetized D-branes in $\T^{2n}$. Finally, Appendix \ref{ap:abe} describes the computation of Abelian flavor-independent discrete gauge symmetries in $\IZ_2 \times \IZ_2'$.

\section{Intersecting branes and the $\IZ_2 \times \IZ_2'$ orbifold}
\label{s:review}

Our examples of D-brane models with discrete flavor symmetries will be based on the toroidal $\IZ_2 \times \IZ_2'$ orientifold background analyzed in \cite{bcms05} (see also \cite{aadds99,dt05,fz08}). As pointed out in there, in this background one can reproduce the main features of realistic D-brane models in Calabi-Yau compactifications, obtaining $\cn = 1$ chiral vacua made up of rigid D-branes. In the following we will briefly review the construction of this class of models, emphasizing those features which are more relevant for the analysis of discrete flavor symmetries. In doing so we will follow the notation and formalism of \cite{bcms05}, which was mainly developed for models of intersecting D6-branes. Nevertheless, such models have a well-known T-dual description in terms of magnetized D-branes, a fact that we will exploit in section \ref{s:magnet} to understand discrete flavor symmetries from an alternative viewpoint closer to the analysis of  \cite{bcmru12}. 

\subsection{Branes at angles in the $\IZ_2 \times \IZ_2'$ orbifold}

\begin{figure}[htb]
  \centering
  \def\svgwidth{440pt}
  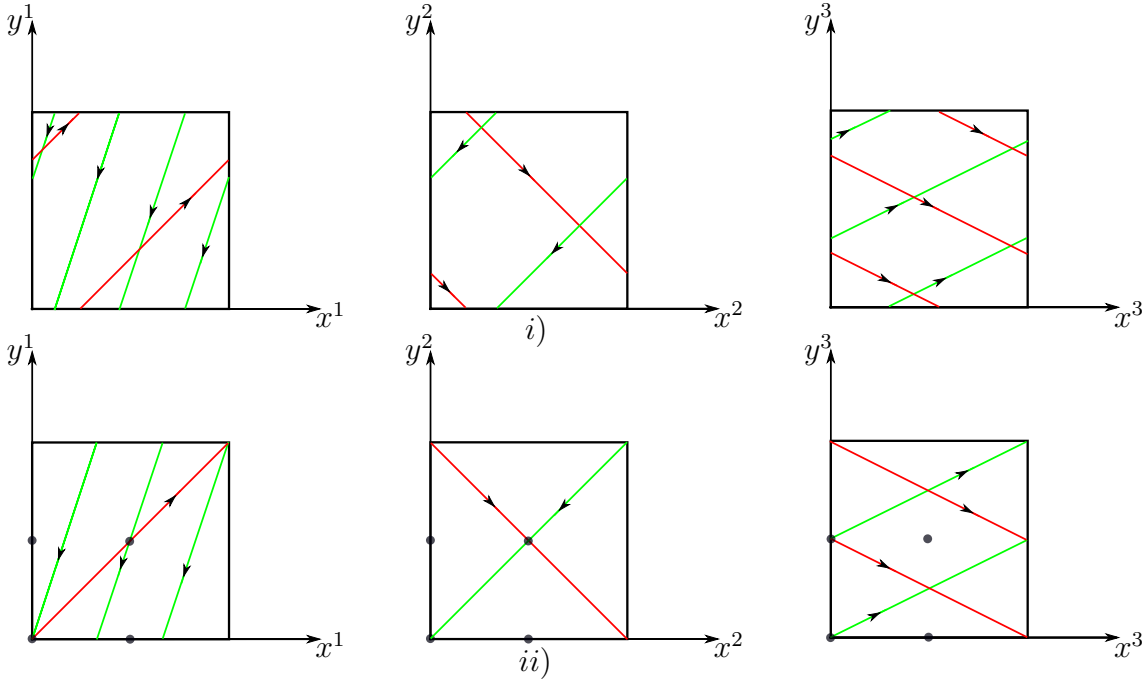
  \caption{Branes with wrapping numbers $(n_a^i, m_a^i) = (1,1)\otimes(1,-1)\otimes(2,-1)$ (red)  and $(n_b^i, m_b^i) = (-1,-3)\otimes(-1,-1)\otimes(2,1)$ (green)  with $i)$ generic values of the position moduli and $ii)$ stuck at the fixed points in the $\IZ_2 \times \IZ_2'$ orbifold. The number of chiral families for $i)$ is $I_{ab}^{\T^6} = (-2)\times (-2)\times 4 = 16$ and for $ii)$ is $I_{ab} = 4$.}
  \label{fig:ab}
\end{figure}

As emphasized in the literature (see \cite{the book,reviews} for reviews on the subject) a quite successful approach to construct particle physics models from string theory is by considering models of intersecting D6-branes in type IIA Calabi-Yau compactifications. The simplest setup in which this approach can be implemented is by taking the compactification space to be a factorized six-torus $\T^6 = (\T^2)_1 \times (\T^2)_2 \times (\T^2)_3$ and adding sets of D6-branes that fill up 4d Minkowski space and wrap different three-dimensional slices of these six extra dimensions. Typically one considers that each D6-brane wraps a product of three one-cycles on the factorized $\T^6$, namely
\be
\label{torcycle}
[\Pi_\a]=\bigotimes_{i=1}^3 \left( n^i_\a\ [a^i] +  m^i_\a\ [b^i] \right) \quad \quad n^i_\a, m^i_\a \in \IZ {\rm \ and \ coprime}
\ee
where $[a^i]$, $[b^i]$ correspond to the two fundamental one-cycles of $(\T^2)_i$. We show in figure \ref{fig:ab}.$i)$ an example of two of these D6-branes $a$ and $b$ wrapping the three-cycles $\Pi_a$ and $\Pi_b$ respectively. If we now wrap $N_a$ D6-branes on top of the three-cycle $\Pi_a$ and $N_b$ on top of $\Pi_b$ we will have a 4d $U(N_a) \times U(N_b)$ gauge group upon dimensional reduction, with a 4d chiral fermion in the $(N_a, \bar{N}_b)$ representation at each intersection point. Hence, by considering several sets of D6-branes one can construct 4d effective theories similar to the Standard Model or extensions thereof \cite{bgkl00,afiru00ph,afiru00,bkl00,imr01}. 

In this setup the relative orientation of a pair of D6-branes is specified in terms of three angles $\theta_{ab}^i$, one per two-torus $(\T^2)_i$. One can render these two D6-branes mutually BPS by applying certain conditions to these angles \cite{bdl96}. However, in order to construct a consistent four-dimensional chiral and $\cn=1$ supersymmetric (and hence stable) model one needs the presence of negative tension objects like O6-planes and to replace the compactification manifold $\T^6$ by a toroidal orbifold of the form $\T^6/\Gamma$, with $\Gamma$ a discrete symmetry group \cite{csu01}. We will leave the effects of adding the O6-planes for section \ref{s:example} and focus here on the implications of having D6-branes at angles in a toroidal orbifold rather than on $\T^6$. 

In particular we will consider type IIA string theory compactified on the background $\T^6/\IZ_2\times \IZ_2'$, the $\IZ_2$  generators acting as
\be
\Theta:
\left\{
\begin{array}{l}
\vspace*{-.25cm}
z_1\to -z_1 \\ \vspace*{-.25cm}
z_2\to -z_2 \\ z_3\to z_3 
\end{array}\right.
\quad\quad\quad\quad
\Theta':
\left\{
\begin{array}{l}
\vspace*{-.25cm}
z_1\to z_1 \\ \vspace*{-.25cm}
z_2\to -z_2 \\ z_3\to -z_3 
\end{array}\right.
\label{orbiaction}
\ee 
on the three complex coordinates of $\T^6 = (\T^2)_1 \times (\T^2)_2 \times (\T^2)_3$. This is the sort of background considered in \cite{bcms05} in order to construct semi-realistic models made of {\it rigid} D6-branes. These rigid or fractional D6-branes wrap three-cycles that are left invariant by (\ref{orbiaction})  and so, unlike in the case of $\T^6$, it is not possible to displace them transversely.\footnote{The fact that fractional D6-branes are fully rigid is a consequence of the choice of discrete torsion in the $\IZ_2\times \IZ_2'$ orbifold. Such choice implies that this orbifold contains collapsed three-cycles at the fixed loci of (\ref{orbiaction}), see appendix \ref{ap:review} and ref.\cite{bcms05} for more details. In a $\IZ_2 \times \IZ_2$ orbifold with the opposite choice of discrete torsion fractional D6-branes are not rigid \cite{csu01}.} As illustrated in figure \ref{fig:ab}.$ii)$ on each $(\T^2)_i$ a fractional D6-brane goes through two fixed points of the action $z_i \to -z_i$. One can determine which are these two fixed points in terms of  the wrapping numbers $(n_a^i, m_a^i)$, as indicated in table \ref{fixed}. From the effective field theory viewpoint the fact that a D6-brane $a$ is rigid implies that at low energies there will be no multiplets in the adjoint representation of $U(N_a)$. One can then build chiral $\CN=1$ models where non-Abelian gauge groups are asymptotically free \cite{bcms05}, a required feature for realistic models in Calabi-Yau compactifications.

\begin{table}[htb]
\renewcommand{\arraystretch}{1.25}
\begin{center}
\begin{tabular}{|c|c|c|}
\hline
$(n^i,m^i)$ & Fixed points on $(\T^2)_i$ \\
\hline \hline
(odd, odd) & $\{1,4\}$ or $\{2,3\}$\\
(odd, even) & $\{1,3\}$ or $\{2,4\}$\\
(even, odd) & $\{1,2\}$ or $\{3,4\}$\\
\hline
\end{tabular}
\caption{\small Fixed points of a 1-cycle on a $\T^2/\IZ_2$ in terms of its wrapping numbers.}
\label{fixed}
\end{center}
\end{table}

\subsection{Chirality and the orbifold projection}
\label{sub:orbichiral}

Let us now consider $N_a$ D6-branes wrapping the three-cycle $\Pi_a$ of $\T^6$ and $N_b$ of them wrapping $\Pi_b$, with wrapping numbers
\be
\begin{array}{lcccc}
\Pi_{a} & : & (n_a^1, m_a^1) &  (n_a^2, m_a^2)  & (n_a^3, m_a^3) \\
\Pi_{b} & : & (n_b^1, m_b^1) & (n_b^2, m_b^2) & (n_b^3, m_b^3) 
\end{array}
\label{D6ab}
\ee
as stated above, this D6-brane sector yields a 4d $U(N_a) \times U(N_b)$ gauge group at low energies, together with $\cn=1$ chiral multiplets in the $(N_a, \bar{N}_b)$ representation, one per each point of intersection of these two three-cycles. The chirality and multiplicity of these multiplets is given respectively by the sign and by the absolute value of the topological intersection number $I_{ab}$, which in the case of $\T^6 = (\T^2)_1 \times (\T^2)_2 \times (\T^2)_3$ is given by 
\be
I_{ab}^{\T^6}  =  I_{ab}^1I_{ab}^2I_{ab}^3 = \prod_{i=1}^3 (n_a^i m_b^i- n_b^im_a^i)
\label{topintT6}
\ee

If we now consider rigid D6-branes in $\T^6/\IZ_2 \times \IZ_2'$ the chiral spectrum will be different, because the $\IZ_2 \times \IZ_2'$ action relates several of the intersection points of the two D6-branes and these will no longer be independent degrees of freedom. In particular, one should project out all those zero modes that are not invariant under the $\IZ_2 \times \IZ_2'$ orbifold action, as we now describe. 

Let us consider the case where $I_{ab}^{\T^6} \neq 0$, so that in the toroidal case we have a net number of chiral fermions in the $ab$ sector. Since the D6-branes wrap BPS and factorizable three-cycles, the massless spectrum in the $ab$ sector is given by $|I_{ab}^{\T^6}|$ 4d chiral fermions in the representation $(N_a, \bar{N}_b)$, whose 4d chirality is given by sign$(I_{ab}^{\T^6})$. In fact, by applying the usual CFT rules for computing the open string spectrum between two intersecting D-branes \cite{bdl96,afiru00}, one can associate to each intersection the piece of 10d massless fermion whose SO(8) weight representation is given by \cite{tesis}
\be
r_{ab}\, =\, (r_0; r_1, r_2, r_3) \, =\, \oh \left(s_1s_2s_3; -s_1, -s_2, -s_3 \right)
\label{rfermion}
\ee
where the first entry indicates the 4d chirality, and the other three correspond to the compact extra dimensions. Here $s_i = {\rm sign} (\vartheta_{ab}^i)$, with $\vartheta_{ab}^i$ the angle of intersection in $(\T^2)_i$ measured in anti-clockwise sense. Notice that $s_i =  {\rm sign} (I_{ab}^i)$ and so 4d chirality is indeed given by ${\rm sign} (I_{ab}^{\T^6})$. In our conventions $I_{ab}^{\T^6}>0$ corresponds to 4d left-handed fermions.

When introducing the orbifold projection, some of these massless fields will be projected out. In particular, we must require that the internal fermionic wavefunctions are invariant under the action of the $\IZ_2 \times \IZ_2'$ generators. These act on a fermion with Lorentz indices (\ref{rfermion}) as
\be
\begin{array}{rl}
\Theta\, : &  \Psi(z_1,z_2,z_3) \ \mapsto \ e^{i \pi (r_1 - r_2)}  \Psi(-z_1,-z_2,z_3) \, =\, s_1s_2 \Psi(-z_1,-z_2,z_3)\\
\Theta'\, : &  \Psi(z_1,z_2,z_3) \ \mapsto \ e^{i \pi (r_2 - r_3)}  \Psi(-z_1,-z_2,z_3) \, =\, s_2s_3 \Psi(z_1,-z_2,-z_3)
\end{array}
\label{orbiferm}
\ee
%
A generic open string wavefunction, which are basically delta functions localized at the intersection points, will not be invariant under such transformations, and so one must form linear combinations that transform appropriately under internal coordinate reversal. 

\begin{figure}[htb]
  \centering
  \def\svgwidth{440pt}
  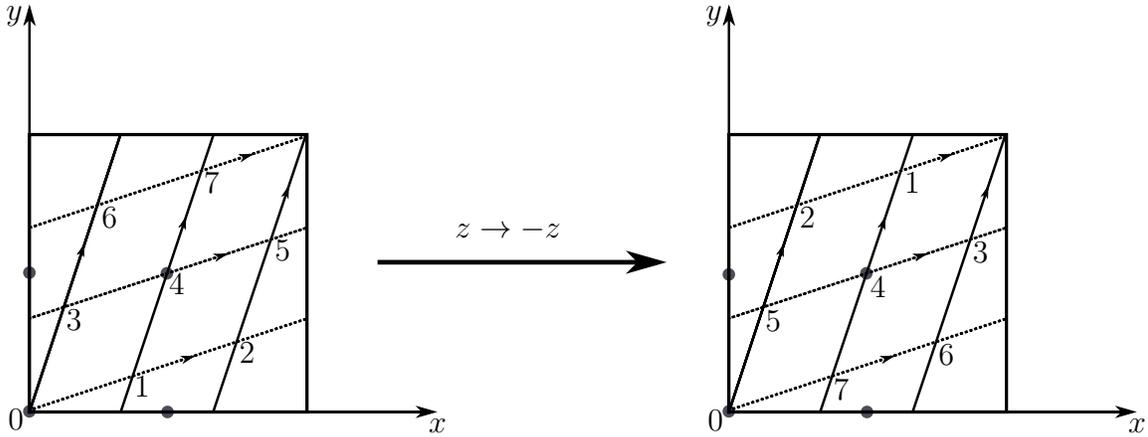
  \caption{Space inversion in one of the tori and D-branes with wrapping numbers $(1,3)$ and $(3,1)$.  Some of the intersection points are invariant while others get exchanged. The even combination of points are $\{0,\,1+7,\,2+6,\,5+3,\,4\}$ and the odd ones $\{1-7,\, 2-6,\,5-3\}$. Notice that the number of even and odd points is in agreement with (\ref{evenodd}).}
  \label{fig:action}
\end{figure}

In order to describe these combination of wavefunctions let us first consider two D-branes wrapping 1-cycles of $(\T^2)_i$ going through the origin, and see how their intersection points transform under the  $\IZ_2$ action generated by $z \mapsto - z$. As shown in figure \ref{fig:action}, one can take linear combinations of delta functions at the intersection points, in order to form wavefunctions which are even and odd under the orbifold action. Such linear combinations have coefficients $\pm1$, and  the number of even and odd points for a given intersection number $I_{ab}^i$ is given by
\be
I_{e}^{i} \, =\, \oh \left(I _{ab}^i + s_i \rho_i \right) \quad \quad I_{o}^{i} \, =\, \oh \left(I _{ab}^i - s_i \rho_i \right)
\label{evenodd}
\ee
where $s_i =  {\rm sign} (I_{ab}^i)$ and 
\be
\rho_i \, \equiv \,
 \left\{
\begin{array}{l}
\vspace*{-.2cm}
1 {\rm \quad for \ \ } I_{ab}^i  {\rm \ \ odd}\\
2 {\rm \quad for \  \ } I_{ab}^i  {\rm \ \ even}
\end{array}\right.
\label{rho}
\ee

Going back to intersecting D6-brane on $(\T^2)_1 \times (\T^2)_2 \times (\T^2)_3 /\IZ_2\times \IZ_2'$, from (\ref{orbiferm}) it is clear that we need to impose that our wavefunctions satisfy
\be
 \Psi(z_1,z_2,z_3)\, =\, s_1s_2 \Psi(-z_1,-z_2,z_3)\, =\, s_2s_3 \Psi(z_1,-z_2,-z_3)
 \label{proj1}
\ee
and so depending on the signs of the two-tori intersection numbers $I_{ab}^i$ we will have to impose different projections. In particular we have that for $I_{ab}^B \neq 0$ and
\begin{itemize}
\item[-] $s_1s_2>0$, $s_2s_3>0$ \quad$\Longrightarrow\quad \Psi = \psi_e^{j_1} \psi_e^{j_2} \psi_e^{j_3}$ \, or \, $  \psi_o^{j_1} \psi_o^{j_2} \psi_o^{j_3}$
\item[-] $s_1s_2>0$, $s_2s_3<0$ \quad$\Longrightarrow\quad \Psi = \psi_e^{j_1} \psi_e^{j_2} \psi_o^{j_3}$ \, or \, $ \psi_o^{j_1} \psi_o^{j_2} \psi_e^{j_3}$
\item[-] $s_1s_2<0$, $s_2s_3>0$ \quad$\Longrightarrow\quad \Psi = \psi_o^{j_1} \psi_e^{j_2} \psi_e^{j_3}$ \,  or \, $  \psi_e^{j_1} \psi_o^{j_2} \psi_o^{j_3}$
\item[-] $s_1s_2<0$, $s_2s_3<0$ \quad$\Longrightarrow\quad \Psi = \psi_e^{j_1} \psi_o^{j_2} \psi_e^{j_3}$ \, or \, $  \psi_o^{j_1} \psi_e^{j_2} \psi_o^{j_3}$
\end{itemize}
where $\psi_e^{j_i}$ runs over even combinations of intersection points on $(\T^2)_i$, and $\psi_o^{j_i}$ is an odd combination of delta-wavefunctions in $(\T^2)_i$. For the first case above we have that the number of generations after the orbifold projection is given by $I_{ab}=I_e^1\,I_e^2\,I_e^3+I_o^1\,I_o^2\,I_o^3$ or 
\be\nonumber
I_{ab}\, =\, \frac{1}{4} \left[I _{ab}^1I _{ab}^2I _{ab}^3 +  s_2s_3\rho_2\rho_3 I _{ab}^1 +  s_1s_3\rho_1\rho_3 I _{ab}^2 +   s_1s_2\rho_1\rho_2 I _{ab}^3 \right] 
\ee
Hence, after imposing that $s_1s_2>0$ and $s_2s_3>0$ we recover the result
\be
I_{ab}  =  \frac{1}{4} \left[ I_{ab}^1I_{ab}^2I_{ab}^3 +   I_{ab}^1\, \rho_2\rho_3  +  I_{ab}^2\, \rho_1\rho_3 +  I_{ab}^3 \, \rho_1\rho_2 \right] 
\label{topint}
\ee
A different choice of signs $s_1$, $s_2$, $s_3$ will select different parities for the wavefunctions of each two-torus, and so a different total number of chiral fermions. Nevertheless, the final expression for $I_{ab}$ will again be given by (\ref{topint}). Notice that this result matches eq.(\ref{ap:topint}), which has been obtained in appendix \ref{ap:review} by means of the topological techniques of \cite{bcms05}.

The same statement holds if we consider the case where the toroidal intersection number $I_{ab}^{\T^6}$ vanishes, as we now briefly discuss. Let us for instance consider the case where only $I_{ab}^1 = 0$.\footnote{The case with two vanishing intersection numbers $I_{ab}^i$ does not correspond to D6-branes preserving $\CN=1$ supersymmetry, and will not be considered here, while the case which all three $I_{ab}^i = 0$ is trivial.} Then instead of (\ref{rfermion}) we have $I_{ab}^2I_{ab}^3$ fermions of the form
\be
r_{ab}\, =\, \oh \left( s_2s_3; - , -s_2, -s_3 \right) \quad {\rm and} \quad \oh \left( -s_2s_3; + , -s_2, -s_3 \right)
\label{rfermion2}
\ee
that is, a non-chiral spectrum. The orbifold action reads
\be
\begin{array}{rl}
\Theta\, : &  \Psi(z_1,z_2,z_3) \ \mapsto \  \pm s_2 \Psi(-z_1,-z_2,z_3)\\
\Theta'\, : &  \Psi(z_1,z_2,z_3) \ \mapsto \  s_2s_3 \Psi(z_1,-z_2,-z_3)
\end{array}
\label{orbiferm2}
\ee
and so we arrive at the following wavefunctions
\begin{itemize}
\item[-] $s_2>0$, $s_3>0$ \quad$\Longrightarrow \quad \Psi = \psi_e^{j_2} \psi_e^{j_3}$ \, or \, $   \psi_o^{j_2} \psi_o^{j_3}$
\quad$\Longrightarrow \quad I_{ab} = I_e^2I_e^3 - I_o^2I_o^3$
\item[-] $s_2>0$, $s_3<0$ \quad$\Longrightarrow \quad \Psi = \psi_e^{j_2} \psi_o^{j_3}$ \, or \, $   \psi_o^{j_2} \psi_e^{j_3}$
\quad$\Longrightarrow \quad I_{ab} = I_e^2I_o^3 - I_o^2I_e^3$
\item[-] $s_2<0$, $s_3>0$ \quad$\Longrightarrow \quad \Psi = \psi_o^{j_2} \psi_e^{j_3}$ \, or \, $   \psi_e^{j_2} \psi_o^{j_3}$
\quad$\Longrightarrow \quad I_{ab} = I_o^2I_e^3 - I_e^2I_o^3$
\item[-] $s_2<0$, $s_3<0$ \quad$\Longrightarrow \quad \Psi = \psi_o^{j_2} \psi_o^{j_3}$ \, or \, $   \psi_e^{j_2} \psi_e^{j_3}$
\quad$\Longrightarrow \quad I_{ab} = I_o^2I_o^3 - I_e^2I_e^3$
\end{itemize}
where the relative minus sign in the expression for $I_{ab}$ comes from the fact that the two fermions in (\ref{rfermion2}) have opposite 4d chirality. Again, in each of the four cases above we find that the index of net chirality $I_{ab}$ matches the expression (\ref{topint}) with $I_{ab}^1 = 0$. 

To summarise, by looking at the action of the orbifold on the open string degrees of freedom we can recover the chiral index obtained in \cite{bcms05} via topological methods. While we have focused on the fermionic modes, the same result is obtained by looking at the light scalars at the D6-brane intersections. The method used here to compute the chiral index $I_{ab}$ is perhaps more involved that the one in \cite{bcms05}, but it also carries more information. First, for the case where $I_{ab}^{\T^6} = 0$ not only does it compute the net chiral index (\ref{topint}) but also detects massless particles of opposite chirality that contribute with opposite signs to the index.\footnote{This spectrum is only computed at tree-level in the string coupling $g_s$, so one expects that vector-like pairs of massless particles will gain a mass by means of quantum corrections, unless some discrete symmetry forbids such mass term. See section \ref{s:example} for an example.} Second, this method not only gives the 4d massless spectrum, but also the explicit expression for the open string wavefunctions in the internal dimensions of the compactification. As we will now see, this will be crucial for studying in detail the discrete flavor symmetries that appear in this class of models.


\section{Discrete flavor symmetries for intersecting branes}
 \label{s:intersec}
 
Having reviewed D-brane models on $\T^6/\IZ_2 \times \IZ_2'$ and how family replication arises for them, we now turn to show the emergence of discrete flavor symmetries in such models. More precisely, we will describe how the Dihedral group $D_4$ and tensor products of it arise in this context, and how the different families transform non-trivially under them. 

As discussed in \cite{bcmru12}, $D_4$ and other non-Abelian discrete flavor symmetries naturally arise in the context of D-brane models, and in particular for models of magnetized D-branes on $\T^{2n}$. There one can detect discrete {\em gauge} flavor symmetries in terms of a 4d effective Lagrangian obtained via dimensional reduction. As shown in appendix \ref{ap:inter} such Lagrangian can also be obtained from models of intersecting D-branes on $\T^{2n}$, and so in principle one can apply the 4d methods of \cite{bcmru12} to detect discrete gauge symmetries. However, it turns out that in models of intersecting D-branes the presence of discrete flavor symmetries can be detected geometrically as well. This is particularly useful to describe them in $\T^6/\IZ_2 \times \IZ_2'$, where applying dimensional reduction is not obvious for certain sectors. In the following we will apply such geometric approach first for a toroidal background and then for $\T^6/\IZ_2 \times \IZ_2'$. Finally, this geometric picture allows to quickly detect when there is an exact discrete flavor symmetry and when such symmetry is just approximate, as we briefly discuss.

 \subsection{Flavor symmetries on the torus}
 
Let us consider type IIA string theory compactified on $\T^6 = (\T^2)_1 \times (\T^2)_2 \times (\T^2)_3$. Such manifold contains 6 continuous isometries $(x_i, y_i) \raw (x_i + \lam_{x_i}, y_i+ \lam_{y_i})$, $i=1,2,3$ which, upon dimensional reduction of the metric, manifest as a $U(1)^6$ gauge group in the 4d effective theory. We will represent such 4d gauge bosons respectively as $V_\mu^{x_i}$ and $V_\mu^{y_i}$.

Let us now introduce a D6-brane wrapping a factorizable three-cycle $\Pi_a$ of the form (\ref{torcycle}). Geometrically, it is clear that the presence of such three-cycle breaks the invariance under translations along the three directions of $\T^6$ transverse to the D6-brane worldvolume, while in the three directions parallel to $\Pi_a$ the translational isometries remain unbroken. From the effective field theory viewpoint three generators of the initial $U(1)^6$ gauge group become massive via a St\"uckelberg mechanism, in which the D6-brane scalars $\phi_a^i$ that parametrize the transverse displacement of $\Pi_a$ in $(\T^2)_i$ are eaten by the generators of the corresponding isometry. Following appendix \ref{ap:inter}, the St\"uckelberg Lagrangian reads
\be\label{st1}
\CL_{\rm St}\, =\, -\frac{1}{2}\sum_{i=1}^3 \left (\p_\mu\phi_a^i-m_a^i V_{\mu}^{x_i} +n_a^i V_{\mu}^{y_i} \right )^2
\ee
 and so the bulk gauge symmetry $U(1)^6$ is broken down to $U(1)^3 \times \IZ_{q_1} \times \IZ_{q_2} \times \IZ_{q_3}$. The $U(1)$ factors are generated by the massless combinations $n_a^i V_{\mu}^{x_i} + m_a^i V_{\mu}^{y_i}$, while the factors $\IZ_{q_i}$ are the discrete remnants of the broken $U(1)$ symmetries generated by $n_a^i V_{\mu}^{y_i} -m_a^i V_{\mu}^{x_i}$. This symmetry breaking pattern is similar to the one studied in \cite{bcmu12}, section 2.5, from where one deduces that $q_i = (n_a^i)^2 + (m_a^i)^2$.
 
Needless to say, adding more D6-branes will further break the translational symmetry. In particular, one would expect that by adding a D6-brane on a three-cycle $\Pi_b$ that intersects $\Pi_a$ transversally all continuous symmetries are broken. Indeed, one then finds that the Lagrangian reads
\be
\CL_{\rm St}\, =\, -\frac{1}{2}\sum_{i=1}^3 \left(\p_\mu\phi_a^i-m_a^i V_{\mu}^{x_i} +n_a^i V_{\mu}^{y_i} \right)^2 + \left(\p_\mu\phi_b^i-m_b^i V_{\mu}^{x_i} +n_b^i V_{\mu}^{y_i} \right)^2
\label{StV2}
\ee
and so if $I_{ab}^i = n_a^im_b^i - n_b^im_a^i \neq 0\ \forall i$ then all gauge bosons $V_\mu^{x_i},V_\mu^{y_i},$ become massive. In fact, as discussed in appendix \ref{ap:inter} the remaining discrete gauge symmetry is given by 
\be
{\bf \ct}^{ab}_{\T^6} \, =\, \IZ_{I_{ab}^1} \times  \IZ_{I_{ab}^2} \times  \IZ_{I_{ab}^3}
\label{disT6}
\ee
Finally, additional D6-branes on three-cycles $\Pi_c$, $\Pi_d$, etc may further break this symmetry. 

\begin{figure}[htb]
  \centering
  \def\svgwidth{350pt}
  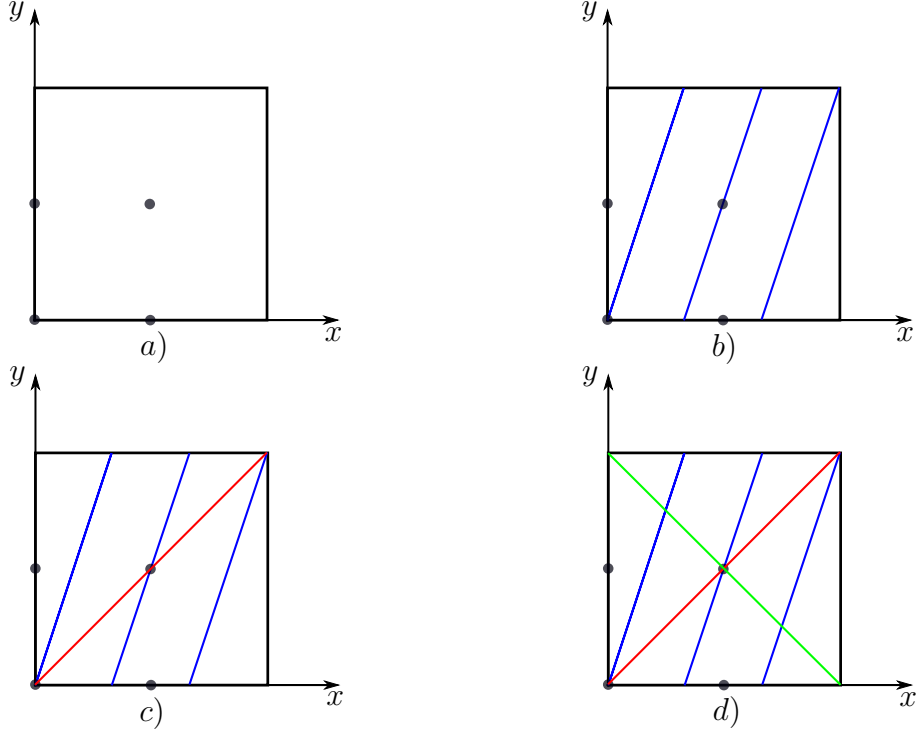
  \caption{$\T^2/\mathbb Z_2$ with a) no branes b) one brane on the cycle (1,3) c) two branes on (1,3) and (1,1) d) three branes on (1,3), (1,1) and (1,-1).}
  \label{fig:4en1}
\end{figure}

While understanding discrete symmetries from the viewpoint of the effective theory is quite powerful, it is quite instructive to develop a more geometrical picture of their meaning. For this let us focus on one of the $\T^2$ factors of $\T^6$. As shown in figure \ref{fig:4en1}a, the absence of D-branes implies a $U(1)^2$ gauge symmetry that corresponds to invariance of the background upon translation in the $x$ and $y$ coordinates. Adding a D-brane $a$ on a 1-cycle $n[a]+ m[b]$ partially breaks this translational symmetry (fig. \ref{fig:4en1}b): infinitesimal translations in the direction $\vec{v}_\| = (n,m)$ leave the geometry invariant while those along $\vec{v}_\perp = (-m,n)$ do not. Nevertheless, finite translations along $\vec{v}_\perp$ do leave the geometry invariant and these, upon quotienting by the coordinate identifications of $\T^2$, generate a discrete group $\IZ_{q}$ with $q = n^2 + m^2$. One then obtains a gauge group $U(1) \times \IZ_q$. Adding a second D-brane $b$ that intersects the first one (fig. \ref{fig:4en1}c) will totally break the invariance under infinitesimal translations. Still, a discrete translational symmetry remains, given by the cyclic permutation of the intersection points of the two D-branes, and this generates a $\IZ_{I_{ab}}$ gauge symmetry. Applying this result to each $(\T^2)_i$ factor of $\T^6$ we obtain (\ref{disT6}).\footnote{In general, given a $\T^{2n}$ geometry we have a $U(1)^{2n}$ translational symmetry. If we introduce two $n$-cycles $\Pi_a^n$, $\Pi_b^n$ that are each a $\T^n \subset \T^{2n}$ and that intersect transversally, then the group of translational symmetry is broken to ${\bf \ct} = \Gamma/\hat{\Gamma}$, where $\Gamma$ is the lattice generated by the intersection points and $\hat{\Gamma}$ is the lattice of coordinate identifications that defines $\T^{2n}$. When $\T^{2n}$ is factorizable $\ct$ is a direct product of discrete subgroups, as in (\ref{disT6}).} From this geometrical perspective one can also see that we are indeed dealing with a flavor symmetry, that acts on the intersection points of each $\T^2$ as the shift generator
\be
g_{\ct} \, =\, 
\left(
\begin{array}{ccccc}
 & 1 \\
 & & 1 \\
 & & & \ddots \\
 & & & & 1 \\
 1 
\end{array}
\right)
\label{clock}
\ee
Finally, this symmetry is further broken if we include additional D-branes (fig. \ref{fig:4en1}d). One can check that if we add a D-brane $c$ then the fundamental region of $\T^2$ will be divided into $d$ identical regions, with $d = {\rm g.c.d.} (I_{ab}, I_{bc}, I_{ca})$ \cite{yukint}. The remaining discrete gauge symmetry is then $\IZ_d$, which corresponds to the common factor $\IZ_{I_{ab}} \cap \IZ_{I_{bc}} \cap \IZ_{I_{ca}}$ of the symmetries for each pair of D-branes. Going back to the case of $\T^6 = (\T^2)_1 \times (\T^2)_2 \times (\T^2)_3$, we conclude that for a system of three D6-branes the translational symmetry is given by
\be
{\bf \ct}^{abc}_{\T^6} \, =\, \IZ_{d_1} \times  \IZ_{d_2} \times  \IZ_{d_3}
\label{disT62}
\ee
with $d_{\, i} = {\rm g.c.d.} (I_{ab}^i, I_{bc}^i, I_{ca}^i)$. This kind of symmetries will constrain the values of the Yukawa couplings of this sector, as pointed out in \cite{yukint,acko09,bcmru12}. 

In fact, the above is not the complete flavor symmetry of the model, as there are further bulk symmetries that are broken by the presence of the D6-branes. Besides the 4d $U(1)^6$ gauge symmetry arising from the metric there will be a 4d $U(1)^6$ gauge symmetry that comes from the B-field, and is generated by the 4d gauge bosons $B_\mu^{x_i}$, $B_\mu^{y_i}$ that arise upon dimensional reduction. From appendix  \ref{ap:inter}, the St\"uckelberg Lagrangian for a single D6-brane reads
\be\label{st2}
\CL_{\rm St}\, =\, -\frac{1}{2}\sum_{i=1}^3 \left (\p_\mu\xi_a^i-n_a^i B_{\mu}^{x_i} - m_a^i B_{\mu}^{y_i} \right )^2
\ee 
with $\xi_a^i$ the Wilson line modulus of the D6-brane on $(\T^2)_i$. This action also has a simple geometrical interpretation, namely that acting with a B-field gauge generators induces a Wilson line on the D6-brane via pull-back on its worldvolume $\Pi_a$. A gauge transformation along $-m_a^i B_{\mu}^{x_i} +n_a^i B_{\mu}^{y_i}$ will have vanishing pull-back and will remain a symmetry of the background, while one along $n_a^i B_{\mu}^{x_i} + m_a^i B_{\mu}^{y_i}$ will be detected by the D6-brane and the corresponding $U(1)$ symmetry will be broken to a discrete subgroup. One can again see that the remaining symmetry is given by $U(1)^3 \times \IZ_{q_1} \times \IZ_{q_2} \times \IZ_{q_3}$.

Adding further D6-branes will generalize this Lagrangian to 
\be
\CL_{\rm St}\, =\, -\frac{1}{2} \sum_\a \sum_{i=1}^3 \left (\p_\mu\xi_\a^i-n_\a^i B_{\mu}^{x_i} - m_\a^i B_{\mu}^{y_i} \right )^2
\label{StB2}
\ee 
with $\a = a, b, c, \dots$. For a system of two D6-branes $a$ and $b$ the symmetry is broken to 
\be
{\bf \cw}^{ab}_{\T^6} \, =\, \IZ_{I_{ab}^1} \times  \IZ_{I_{ab}^2} \times  \IZ_{I_{ab}^3}
\label{disT6W}
\ee
 which in principle looks similar to (\ref{disT6}) but the action of the generators on the flavor degrees of freedom is quite different. In this case the generator of the flavor symmetry acts on the intersection points of each $\T^2$ as the clock generator\footnote{The action of $B_\mu$ is equivalent to switching on a Wilson line, $A^\a = d\chi^\a = 2\pi d \zeta^\a$, where $\zeta^\a \sim \zeta^\a + 1$ is the coordinate of the D6-brane $\a$ along the corresponding $\T^2$. An open string located at $\zeta^\a = j/N$ and with charge $q_\a$ will have its phase shifted as $e^{iq_\a \chi^\a} = e^{2\pi i q_\a j/N}$, from where the action (\ref{shift}) follows.\label{footshift}}
 \be
 g_{\cw} \, =\, 
\left(
\begin{array}{ccccc}
 1 &  \\
 & e^{2\pi i \frac{1}{N}} &  \\
  & & e^{2\pi i \frac{2}{N}}\\
 & & & \ddots \\
 & & & &  e^{2\pi i \frac{N-1}{N}}
\end{array}
\right)
\label{shift}
\ee
with $N = I_{ab}^i$. As it is easy to check the generators (\ref{clock}) and (\ref{shift}) do not commute, and so with their combined action they end up generating the discrete non-Abelian group of the form $H_N \simeq ( \IZ_N \times \IZ_N ) \rtimes \IZ_N$ for each $\T^2$, or more precisely
\be
{\bf P}^{ab}_{\T^6} \, =\, H_{I_{ab}^1} \times  H_{I_{ab}^2} \times  H_{I_{ab}^3}
\label{disT6P}
\ee
which is the result obtained in the T-dual picture of magnetized D9-branes \cite{acko09,bcmru12}. Finally, for a triplet of D6-branes this symmetry is reduced to 
\be
{\bf P}^{abc}_{\T^6} \, =\, H_{d_1} \times  H_{d_2} \times  H_{d_3}
\label{disT6P2}
\ee
with again $d_{\, i} = {\rm g.c.d.} (I_{ab}^i, I_{bc}^i, I_{ca}^i)$
 
\subsection{Flavor symmetries on $\IZ_2 \times \IZ_2'$}

Let us now consider the case of type IIA string theory compactified on $\T^6/\IZ_2 \times \IZ_2'$. Unlike the case of $\T^6$ the orbifold background does not have any continuous isometry even in the absence of D-branes. Hence the dimensional reduction that led to effective actions of the form (\ref{StV2}) or (\ref{StB2}) does not apply, and we need to use a different method to determine which are the discrete flavor symmetries that can arise in this case. Notice that the same will be true in Calabi-Yau compactifications, as these manifolds do not contain any continuous isometry either. 

Fortunately in our discussion of $\T^6$ we have developed an alternative method for detecting discrete flavor symmetries. For instance, in the case of translational isometries the flavor symmetry was understood as the group of isometries of the manifold that is also preserved by the D-brane configuration. This observation can also be applied to the $\T^6/\IZ_2 \times \IZ_2'$ orbifold, whose group of translational isometries is discrete and given by $\IZ_2^{\, 6}$. Upon dimensional reduction this will give rise to a 4d discrete $\IZ_2^{\, 6}$ gauge group that will be broken to a subgroup by the inclusion of D6-branes, and this subgroup will be part of the discrete flavor symmetry of the model. 

To get an idea of this symmetry breaking let us again consider the toy example $\T^2/\IZ_2$. The $\IZ_2$ quotient is generated by $z \mapsto -z$ and so there are four fixed points that break the continuous isometry group $U(1)^2$ of $\T^2$ down to $\IZ_2 \times \IZ_2$. The generators of this discrete group are the actions $z \mapsto z + 1/2$ and $z \mapsto z + \tau/2$, with $\tau$ the complex structure of the torus, that interchange the fixed points at $\{0, 1/2, \tau/2, (1+\tau)/2\}$ among them.

 \begin{figure}[htb]
  \centering
  \def\svgwidth{650pt}
  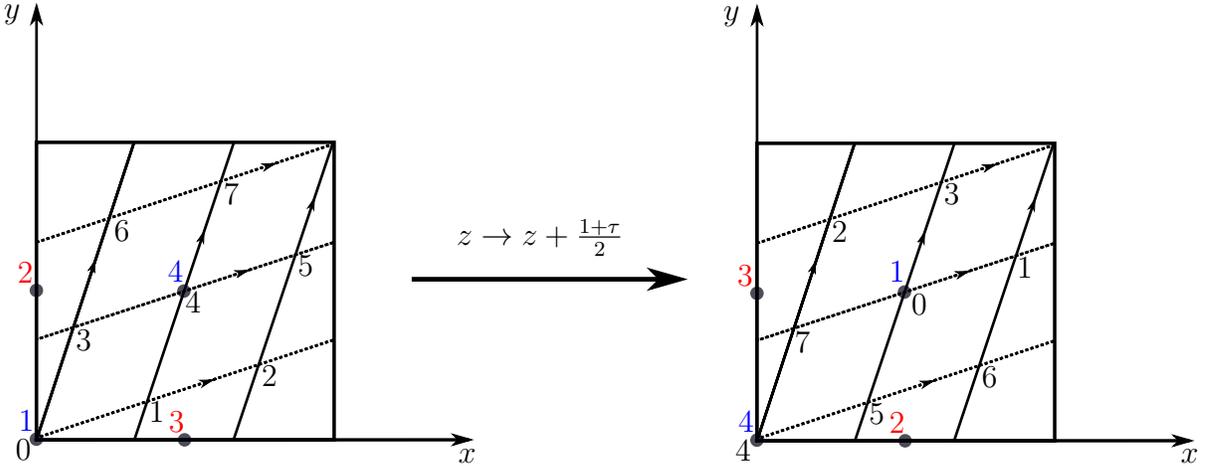
  \caption{Translation $z\rightarrow z+\frac{1+\tau}{2}$ in a square torus. This is the generator of the shift symmetry in the intersecting brane picture.}
  \label{fig:clock}
\end{figure}

Let us now introduce D-branes in this background. In our toy example fractional D-branes are represented by 1-cycles that pass through two of the four fixed points of $\T^2/\IZ_2$. It is then clear that the presence of a single D-brane breaks the group of translations $\IZ_2 \times \IZ_2$ down to $\IZ_2$, where this latter $\Z_2$ interchanges the two fixed points that the D-brane goes through. For instance, as shown in figure \ref{fig:clock} a D-brane whose wrapping numbers $(n,m)$ are both odd will go through the fixed points \{1,4\} or \{2,3\} (see table \ref{fixed}). The symmetry of this system is then the $\IZ_2$ generated by $z\rightarrow z+\frac{1+\tau}{2}$ that interchanges the fixed points as $1 \leftrightarrow 4$ and $2 \leftrightarrow 3$. As figure \ref{fig:clock} also shows this $\Z_2$ symmetry will still be preserved after we introduce a second D-brane, provided that it also goes through the fixed points \{1,4\} or \{2,3\} or, in other words, if its wrapping numbers $(n,m)$ are both odd as well. In general, a pair of fractional 1-cycles on $\T^2/\IZ_2$ will preserve a $\IZ_2$ translational symmetry if they belong to the same row of table \ref{fixed}, which is equivalent to asking that the intersection number $I_{ab} = n_am_b - n_bm_a$ is even. Finally, three or more 1-cycles will preserve the same $\Z_2$ symmetry if they all belong to the same row of table \ref{fixed}, or in other words if all the pairwise intersection numbers $I_{ab}, I_{bc}, I_{ca},\dots$ are even. 

One may now generalize these observations to the case of $(\T^2)_1 \times (\T^2)_2 \times (\T^2)_3/\IZ_2 \times \IZ_2'$, as each $(\T^2)_i$ factor will behave like our toy example. Instead of our previous result (\ref{disT6}) for a pair of D6-brane on $\T^6$ we now have that each $(\T^2)_i$ factor contributes at most with a $\IZ_2$ symmetry, and only if the intersection number $I_{ab}^i$ is even. Hence
\be
{\bf \ct}^{ab}_{\T^6/\IZ_2 \times \IZ_2'} \, =\, \IZ_{\rho_1} \times  \IZ_{\rho_2} \times  \IZ_{\rho_3}
\label{disZ2}
\ee
with $\rho_i$ defined as in (\ref{rho}). Similarly, for a system of three D6-branes we have
\be
{\bf \ct}^{abc}_{\T^6/\IZ_2 \times \IZ_2'} \, =\, \IZ_{d_1} \times  \IZ_{d_2} \times  \IZ_{d_3}
\label{disZ22}
\ee
where now $d_{\, i} = {\rm g.c.d.} (2,I_{ab}^i, I_{bc}^i, I_{ca}^i)$. 

Just like in the case of $\T^6$, this will not be the whole flavor symmetry group. There will also be a symmetry group generated by discrete gauge transformations of the B-field, whose gauge bosons $B_\mu^{x_i,y_i}$ are projected out infinitesimally by the orbifold. Just like for finite translations, these finite B-field transformations generate a $\IZ_2^{\, 6}$ group on $\T^6/\IZ_2 \times \IZ_2'$ that is broken to a subgroup when D6-branes are introduced. One can check that this subgroup ${\bf \cw}^{ab}_{\T^6/\IZ_2 \times \IZ_2'}$ is isomorphic to (\ref{disZ2}) when two D6-branes are introduced, and similarly for  ${\bf \cw}^{abc}_{\T^6/\IZ_2 \times \IZ_2'}$ and (\ref{disZ22}) for a triplet of D6-branes. As in the $\T^6$ case the two $\IZ_2$ subgroups that arise from $(\T^2)_i$ do not commute, but rather generate the non-Abelian group $(\IZ_2 \times \IZ_2) \rtimes \IZ_2 \simeq H_2$, which is nothing but the Dihedral group $D_4$. The final symmetry group for a pair of D6-branes would then be given by 
\be
{\bf P}^{ab}_{\T^6/\IZ_2 \times \IZ_2'} \, =\, H_{\rho_1} \times  H_{\rho_2} \times  H_{\rho_3} \, =\, D_4^{[\rho_1-1]} \times  D_4^{[\rho_2-1]} \times D_4^{[\rho_3-1]}
\label{disZ2P}
\ee
where $D_4^{[0]}$ is the trivial group and $D_4^{[1]} = D_4$, while for a D6-brane triplet we should have
\be
{\bf P}^{abc}_{\T^6/\IZ_2 \times \IZ_2'} \, =\, D_4^{[d_1-1]} \times  D_4^{[d_2-1]} \times D_4^{[d_3-1]}
\label{disZ2P2}
\ee
with $d_{\, i} = {\rm g.c.d.} (2,I_{ab}^i, I_{bc}^i, I_{ca}^i)$. 

We will rederive this result in the next section, where we will use the T-dual framework of magnetized D-branes to obtain (\ref{disZ2P}) and (\ref{disZ2P2}), as well as to make contact with the results of \cite{acko09,ako08,acko08,acko09b}. In addition to deriving the symmetry group we will use the magnetized picture to classify under which representations do the chiral families transform on each model. The reader not interested in such details may find the results summarized below, and may safely skip to section \ref{s:example} where they are applied to specific examples.

\subsubsection*{Summary}

In general we find that the wavefunctions that correspond to bifundamental fields $(N_a, \bar{N}_b)$ are of the form 
\be
\psi_{ab}^{j_1, j_2, j_3}\, =\, \psi_{ab}^{j_1} \cdot \psi_{ab}^{j_2} \cdot \psi_{ab}^{j_3}
\ee
where $\psi_{ab}^{j_i}$ lives in $(\T^2)_i$. Depending on the signs of the intersection numbers $I_{ab}^i$ these wavefunctions will be even or odd under the action $z_i \mapsto - z_i$, as discussed below (\ref{proj1}). If $\psi_{ab}^{j_i}$ is even the index $j_i$ will run over $I^i_e$ values and if it is odd over $I_o^i$ values, with $I_e^i$ and $I_o^i$ defined in (\ref{evenodd}). Moreover, if $I_{ab}^i$ is even, this index will transform under a specific representation of the flavor symmetry group $D_4$ of $(\T^2)_i$. This representation will depend on the value of $I_{ab}^i$ and the wavefunction parity, as shown in table \ref{tab:reps}.
\begin{table}[htb] 
\renewcommand{\arraystretch}{1.25}
 \begin{center}
    \begin{tabular}{ | c || c | c |}
    \hline
  $|I_{ab}^i|$ & $\psi^{j_i}_{\rm even}$ \quad \footnotesize{${\rm dim} = |I_{ab}^i|/2 + 1$}  & $\psi^{j_i}_{\rm odd}$ \quad \footnotesize{${\rm dim} = |I_{ab}^i|/2 - 1$} \\ \hline\hline
 {\footnotesize $4s+2$} &  $ \stackrel{s+1}{\oplus}  \mathbf R_2$ & $ \stackrel{s}{\oplus}  \mathbf R_2$\\ \hline
{\footnotesize $8s+4$} & $\stackrel{s+1}{\oplus} (+,+) \stackrel{s+1}{\oplus} (+,-) \stackrel{s+1}{\oplus} (-,+) \stackrel{s}{\oplus} (-,-)$ & $\stackrel{s}{\oplus} (+,+) \stackrel{s}{\oplus} (+,-) \stackrel{s}{\oplus} (-,+) \stackrel{s+1}{\oplus} (-,-)$   \\ \hline
   {\footnotesize $8s+8$} & $\stackrel{s+2}{\oplus} (+,+) \stackrel{s+1}{\oplus} (+,-) \stackrel{s+1}{\oplus} (-,+) \stackrel{s+1}{\oplus} (-,-)$  & $\stackrel{s}{\oplus} (+,+) \stackrel{s+1}{\oplus} (+,-) \stackrel{s+1}{\oplus} (-,+) \stackrel{s+1}{\oplus} (-,-)$   \\ \hline
    \end{tabular}
\end{center}
\caption{Different family representations under the flavor symmetry group $D_4$ on each $(\T^2)_i$.  Here $\mathbf R_2$ stands for the 2-dimensional irreducible representation of the dihedral group $D_4$, see (\ref{R2}), while $(\pm,\pm')$ stands for the one-dimensional representation in which the two generators of $D_4$ act respectively as $\pm \Id$ and $\pm' \Id$.}
\label{tab:reps}
\end{table}

\subsection{Exact versus approximate symmetries}

The previous discussion is quite useful in order to draw a notion of exact and approximate discrete symmetry for this class of models. By exact symmetry it is meant  a discrete {\em gauge} symmetry of the 4d effective field theory, in the sense of \cite{Alford:1988sj,Krauss:1988zc,Alford:1989ch,Preskill:1990bm,Alford:1990mk,Alford:1990pt,Alford:1991vr,Alford:1992yx}. The non-Abelian discrete symmetries discussed in \cite{bcmru12} are of this sort, 
the procedure to detect the gauge nature of a discrete symmetry being the construction of the effective 4d Lagrangian. For the case of the $\IZ_2 \times \IZ_2'$ orbifold, the construction of such 4d Lagrangian  is beyond the scope of this paper, and so we will instead adopt a different approach and discuss the exactness of a discrete flavor symmetry by means of the geometric intuition developed above. 

As pointed out in the last subsection the $\IZ_2 \times \IZ_2'$ orbifold background has a group of translational isometries given by $\ct_{\IZ_2 \times \IZ_2'} = \IZ_2^{\, 6}$. Coming from isometries of the internal manifold, this group naturally translates as a $\IZ_2^{\, 6}$ gauge group in the 4d effective theory. Adding a fractional D6-brane will break this group down to $\IZ_2^{\, 3}$, where each $\IZ_2$ factor comes from a different $(\T^2)_i$. This $\IZ_2^{\, 3}$ symmetry group can be understood as the group of translations that leaves both the orbifold background and the D-brane invariant, and so it is a natural candidate for a 4d discrete gauge symmetry of the orbifold plus D-brane background. The question is now if the whole set of D-branes in a given model will respect such symmetry as well, or in other words if the whole orbifold plus D-brane backgrounds will be invariant under this $\IZ_2^{\, 3}$ translational symmetry or a subgroup thereof.\footnote{Another important element of a D-brane model is the orientifold planes or O-planes, which we have so far ignored. One can check that their presence does not further break these discrete symmetries, at least for the class of models with  rectangular $(\T^2)_i$ that we will consider in section \ref{s:example}.}

By looking at figure \ref{fig:clock} it is clear that a group of two or more D-branes will be invariant under a $\IZ_2$ shift symmetry of $(\T^2)_i$ if all of them go through the same pair of fixed points. In fact, the condition for the symmetry to be exact is weaker, and we only need to require that all the intersection numbers $I_{\a\b}^i$ between D-branes in this two-torus are even. The same is true for the $\IZ_2$ discrete symmetry that arises from the B-field, and so for the whole $D_4$ that both $\IZ_2$ actions generate. Let us denote as $D_4^{(i)}$ the dihedral flavor symmetry that may arise from $(\T^2)_i$, we then have that
\be
D_4^{(i)} {\rm \ is \, exact\ } \iff \ I_{\a\b}^{(i)}\ {\rm \ is\, even\ } \forall \a,\b
\label{cexact}
\ee
and so the group (\ref{disZ2P2}) may become a gauge or exact discrete gauge symmetry of the 4d effective theory depending on the whole set of D-brane intersection numbers.\footnote{Strictly speaking if (\ref{cexact}) is true then $D_4^{(i)}$ is a global symmetry of the 2d action of the BCFT theory, which becomes a local symmetry in target space to all orders in string perturbation theory. One should still check that this symmetry is preserved at the non-perturbative level in the string coupling, something that here is assumed. We leave a more detailed analysis of this subject for future work.} 

It may seem that the exactness condition (\ref{cexact}) is kind of restrictive when constructing explicit D-brane models. However, as follows from the results of \cite{bcms05}, one typically needs that D6-branes go through the same fixed points in order to satisfy the RR twisted tadpole conditions necessary to construct an anomaly-free consistent model. From this viewpoint, the stringy consistency conditions of the model render natural the appearance of exact discrete flavor symmetries in the low energy theory, as the examples of section \ref{s:example} illustrate. 

To be more precise let us consider a $\IZ_2 \times \IZ_2'$ model with $K$ stacks of fractional D6-branes, and let us separate them in two subgroups $A \, =\, \{a_1, a_2, a_3, \dots\}$ and $B\, =\, \{b_1, b_2, \dots \}$, wrapping three-cycles $\Pi_{a_i}, \Pi_{b_j}$ of the orbifold. The group $A$  will yield a 4d gauge group $\prod_i U(N_{a_i})$, whose chiral spectrum will be specified by the intersection numbers $I_{a_ka_l}$. Typically, demanding that this sector of the theory is free of chiral anomalies by itself will impose cancellation of RR twisted tadpoles within the group $A$ of D6-branes, and this will most likely happen when all the D6-branes in $A$ respect the same $D_4^{(i)}$ symmetry in $(\T^2)_i$. There will be then a discrete symmetry group of the form (\ref{disZ2P2}) acting on this sector and constraining its couplings in the 4d effective field theory. 

We may in particular consider the case where the group $A$ of D6-branes contains the spectrum of the Standard Model or an extension thereof, while the group $B$ of additional D6-branes contains an extra (hopefully hidden) sector of the theory. If all the $(\T^2)_i$ intersection numbers $I_{a_ka_l}^i$ are even there will be a flavor symmetry group $D_4^{(i)}$ acting on the visible sector of the theory. However, the extra sector $B$ may not respect such symmetry, and if there is a single intersection number $I_{a_kb_l}^i$ which is odd then $D_4^{(i)}$ will not be an exact symmetry of the model. Nevertheless, it can still be considered an {\em approximate} symmetry of the sector $A$, because all the couplings within this sector must still respect this symmetry at least at tree-level. In particular, in a supersymmetric model the holomorphic Yukawa couplings of this sector will be constrained by the flavor symmetry at all orders in perturbation theory. The K\"ahler potential, on the other hand, may already get symmetry-breaking corrections at the perturbative level by effects involving the D-branes $b_l$ that do not respect the symmetry (e.g., massive open string attached to $b_l$ running in loops). It would be interesting to see if the scenarios and  techniques that apply to approximate continuous symmetries, see e.g. \cite{bchkmq08,maha11}, could also be at work for this case. We leave for the future to explore the phenomenological consequences of these approximate symmetries in realistic and semi-realistic D-brane models. 


\section{Discrete flavor symmetries for magnetized branes}
 \label{s:magnet}

An interesting feature of the D-brane models analyzed in the previous sections is that they have a well-known T-dual description in terms of magnetized D-branes, which is a fruitful arena for understanding flavor symmetries. Indeed, as emphasized in \cite{magnus}, the framework of magnetized D-branes allows to compute 4d effective couplings by first solving for the chiral modes internal wavefunction profile, and then calculating their overlap over the extra dimensions of the  compactification. As pointed out in \cite{acko09}, by inspection of these zero mode wavefunctions one can understand the flavor symmetries present in the model. Finally, it was shown in \cite{bcmru12} how to obtain from this framework a 4d effective Lagrangian describing such discrete gauge flavor symmetries. 

In the following we will rederive our previous results in the dual context of magnetized D-brane models, both in $\T^6$ and in the $\IZ_2 \times \IZ_2$ orbifold. This will allow to perform a more systematic analysis of the discrete flavor symmetries, and in particular to see under which representation transform the different families of chiral multiplets.\footnote{Although more systematic, the field theory framework of magnetized D9-branes is less general than the framework of intersecting D6-branes, because it fails to capture the actual 4d effective theory when the magnetic fluxes are not diluted and/or when anti-D9-branes or D-branes of lower dimension are present. In this sense, the results of this section can be seen as complementary to the ones obtained previously in the context of intersecting D6-branes.}

\subsection{Non-Abelian flavor symmetries from magnetization}

Let us first consider type IIB string theory compactified on the factorized six-torus $\T^6 = (\T^2)_1 \times (\T^2)_2 \times (\T^2)_3$ and $N$ D9-branes filling the whole of 10d space-time. We may add an non-trivial magnetization $\langle F_2 \rangle$ along the coordinates of $\T^6$  without breaking 4d Poincar\'e invariance. In particular, we may choose a $U(N)$ Yang-Mills field strength of the form
\begin{equation}
\label{flux2}
F_2=\sum_{i=1}^3\frac{\pi i}{\textrm{Im }\tau^i}\begin{pmatrix}
\frac{m_a^i}{n_a^i}\mathbb{I}_{N_a} & & & \\
& \frac{m_b^i}{n_b^i}\mathbb{I}_{N_b} & & \\
& & \frac{m_c^i}{n_c^i}\mathbb{I}_{N_c} & \\
& & & \ddots
\end{pmatrix}dz^i\wedge d\bar z^i
\end{equation}
where $z^i=dx^i+\tau^idy^i$ is the complexified coordinate of $(\T^2)_i$,  $N_\a = n^1_\alpha n^2_\alpha n^3_\alpha$, $N = \sum_\a N_\a$. Each block within (\ref{flux2}) can be seen as a different D9-brane with `magnetic numbers' $n_\a^i, m_\a^i \in  \IZ$ and with gauge group $U(d_\a^1d_\a^2d_\a^3)$, $d_\a^{\, i} = {\rm g.c.d.}(n_\a^i, m_\a^i)$ \cite{torons}. One can then describe a pair of D9-branes in terms of these magnetic numbers
\be
\begin{array}{lcccc}
D9_{a} & : & (n_a^1, m_a^1) &  (n_a^2, m_a^2)  & (n_a^3, m_a^3) \\
D9_{b} & : & (n_b^1, m_b^1) & (n_b^2, m_b^2) & (n_b^3, m_b^3) 
\end{array}
\label{D9ab}
\ee
in a rather analogous fashion to (\ref{D6ab}). In fact, both configurations are mapped to each other by performing three T-dualities, as have been used extensively in the literature. In this correspondence, the matter localized at the D6-brane intersections $\Pi_a \cap \Pi_b$ is mapped to the set of zero modes that arise from a $N_a \times N_b$ submatrix of the 10d U(N) adjoint fields $(\Psi, A_M)$ \cite{magnus}. In the following we will assume that $d_\a^{\, i} = {\rm g.c.d.}(n_\a^i, m_\a^i) = 1$ and, in particular, that $n_\a^i =1$ $\forall \a, i$.\footnote{While more involved, one can  generalize the analysis for the case $n_\a^i > 1$, along the lines of \cite{magnus,acko10}. The case where some of the $n_\a^i = 0$ also makes sense, and describes a model with D7, D5 or D3-branes. This case, however, it is difficult to analyze from the field theory viewpoint and it is then more convenient to analyse the discrete flavor symmetries from the T-dual framework of intersecting D6-branes.} This greatly simplifies the analysis, since then $N_\a = 1$ and the internal profile of the 4d chiral zero modes is an scalar wavefunctions $\psi^j$ instead of a matrix of wavefunctions. 

As pointed out in \cite{acko09}, by inspection of such zero mode wavefunctions one can guess the flavor symmetry of a model of  magnetized D-branes. However, as we will now show, one can directly characterize this flavor symmetry group by looking at the symmetries of the D-brane configuration, without solving for any zero mode. For simplicity, let us first consider a $\T^2$ and a $U(2)$ gauge sector with a magnetization
\be
\label{fluxab}
F_2=\frac{\pi i}{\textrm{Im }\tau}\begin{pmatrix}
m_a & \\
& m_b
\end{pmatrix}dz \wedge d\bar z
\ee
that breaks the gauge symmetry down to $U(1)_a \times U(1)_b$. In general this system is interpreted as two magnetized D-branes $a$ and $b$, whose zero modes in the $ab$ sector feel the relative flux $M = m_a - m_b = - I_{ab}^{\T^2}$. Even if $F_2$ is constant we have a vector potential of the form
\be
\label{potab}
A(x,y)=\pi  
\begin{pmatrix}
m_a & \\
& m_b
\end{pmatrix} (xdy - ydx)
\ee
which breaks the invariance under translations. More precisely we have that
\be
\begin{array}{lcr}
A(x+\lam_x,y) & = & 
A(x,y) +  \pi \lam_x \,
(m_a X_a +m_b X_b) dy, \\
A(x, y+\lam_y) & = & 
A(x,y) -  \pi \lam_y \,
(m_a X_a +m_b X_b) d x
\end{array}
\label{transfA}
\ee
where
\be
X_a = 
\left(
\begin{array}{cc}
1 & \\
& 0
\end{array}
\right)
\quad {\rm and} \quad
X_b = 
\left(
\begin{array}{cc}
0 & \\
& 1
\end{array}
\right)
\ee
 Eq.(\ref{transfA}) can be interpreted as the fact that, in the presence of the background flux (\ref{fluxab}), an arbitrary translation is no longer a symmetry of the theory because $\langle A \rangle$ is not invariant under it. Nevertheless, from (\ref{transfA})  we see that this variation is equivalent to a linear gauge transformation, which can in turn be interpreted as a Wilson line. For certain discrete choices of $\lam_x$, $\lam_y$ such Wilson line will be trivial, and this will correspond to a discrete symmetry of the configuration. 
 
To properly see this point let us replace the gauge potential $A$ by a gauge covariant object such as the covariant derivative $i D$. In addition, we must take into account that in a gauge theory translations of the form $x \raw x +\lam_x$ are generated by exp$(\lam_x D_x)$. The gauge covariant version of (\ref{transfA}) is then 
\be
e^{\lam_jD_j} iD_ke^{-\lam_jD_j}=i D_k +  \lam_j F_{jk}
\label{transfDT}
\ee
where $j,k = x,y$, and we have used that $[D_j,D_k]=-iF_{jk}$. In fact, translations are not the only possible gauge transformations that we can perform but, just like in the case of D-branes at angles, there are also the ones generated by the 4d gauge bosons $B_\mu^{x,y}$ that arise from the B-field. These act on the covariant derivative as a diagonal linear gauge transformation, namely
\be
e^{\mu_j B_j } i D_ke^{-\mu_j B_j }=i D_k + \mu_j \d_{jk} \Id_2, \qquad B_x = 2\pi i\, x\, \Id_2, \  B_y = 2\pi i \, y\, \Id_2 
\label{transfDB}
\ee
 Finally, we can write both (\ref{transfDT}) and (\ref{transfDB}) in the form
 \be
 \label{invD}
 \begin{array}{lcr}
e^{\lam_x D_x +\mu_y B_y} \, iD \, e^{-\lam_x D_x -\mu_y B_y}\, =\, \Xi_y\, iD\, \Xi_y^{-1},  \qquad \Xi_y = e^{2\pi i (\xi_{y,a} X_a +\xi_{y,b} X_b)y}\\
e^{\lam_y D_y +\mu_x B_x}\, iD\,  e^{-\lam_y D_y -\mu_x B_x}\, =\, \Xi_x\, iD\, \Xi_x^{-1},  \qquad \Xi_x = e^{2\pi i (\xi_{x,a} X_a +\xi_{x,b} X_b)x}
 \end{array}
 \ee
where
\bea
\label{relWL1}
\xi_{y,a}  = {\lam_x} {m_a} + \mu_y  & \qquad &  \xi_{y,b}  = \lam_x {m_b} + \mu_y \\
 \xi_{x,a}  = - {\lam_y} {m_a} + \mu_x  & \qquad & \xi_{x,b}  =  - \lam_y {m_b} + \mu_x
\label{relWL2}
\eea
with $\xi_{x,\a}$ representing a Wilson line for the gauge group $U(1)_\a$ along the coordinate $x$, and similarly for $\xi_{y,\a}$. Notice that no Wilson lines are induced for $U(1)_a$ if $\mu_y = -{\lam_x} {m_a}$ and $\mu_x = {\lam_y} {m_a}$, and so this gauge sector remains invariant under this particular combined action of the bulk gauge transformations. In other words, the magnetized D-brane $\a=a$ breaks the original $U(1)^4$ symmetry of the bulk down to $(U(1) \times \IZ_q)^2$, similarly to the previous case of a D-brane wrapping a 1-cycle. A similar statement can be made for the D-brane $b$, and it can all be encoded in the  4d effective field theory via the following St\"uckelberg Lagrangian
\be\label{Stumag}
\CL_{\rm St}\, =\, -\frac{1}{2}\sum_{\a=a,b}\left \{ \left (\p_\mu\xi_{x,\a} +m_\a V_{\mu}^{y} - B_{\mu}^{x} \right )^2+ \left (\p_\mu\xi_{y,\a}-{m_\a} V_{\mu}^{x} - B_{\mu}^{y} \right )^2 \right \}
\ee
which is a particular case of (\ref{d9s}), derived in appendix \ref{ap:inter} from dimensional reduction. Here $\xi_{x,\a}$, $\xi_{y,\a}$ represent the 4d scalar fluctuations corresponding to the Wilson lines of $U(1)_\a$, $V_\mu^{x,y}$ are the gauge bosons that arise from the metric and $B_\mu^{x,y}$ from the B-field.

One can now interpret (\ref{invD}) as advanced before: whenever $(\xi_{y,a}, \xi_{y,b}) \in \IZ^2$ we have a trivial Wilson line shift in the rhs of (\ref{invD}), and so the corresponding gauge transformation generated by a $V_\mu^x$ and $B_\mu^y$ is a symmetry of the system. One can check that there are $M = m_a - m_b$ inequivalent values of $(\lam_x, \mu_y)$ that correspond to $(\xi_{y,a}, \xi_{y,b}) \in \IZ^2$, and that such values generate a residual $\IZ_M$  symmetry. Similarly, there are $M$ values of $(\lam_y, \mu_x)$ such that $(\xi_{x,a}, \xi_{x,b}) \in \IZ^2$, and these generate an additional $\IZ_M$ symmetry. Finally, because of (\ref{transfDT}) these two $\IZ_M$ symmetries do not commute, and we end up with a non-Abelian symmetry group given by $H_M \simeq (\IZ_M \times \IZ_M) \rtimes \IZ_M$. 

It is instructive to apply this discrete symmetry to the chiral zero mode wavefunctions of this magnetized system, and in particular to those in the bifundamental representation $(+1,-1)$ of $U(1)_a \times U(1)_b$, which is where families of chiral matter arise from. On these modes $B_{x,y}$ act trivially, so the above discrete symmetry is implemented by\footnote{In \cite{bcmru12} the alternative set of operators was considered
\bea
\nonumber
e^{\frac{n_x}{M} X_x} & \qquad & X_x\, =\, \p_x - \pi i \left(m_a X_a + m_b X_b \right) y\\
e^{\frac{n_y}{M} X_y} & \qquad & X_y\, =\, \p_y + \pi i \left(m_a X_a + m_b X_b \right) x
\nonumber
\eea
in order to implement the action of the flavor symmetry group on wavefunctions. Both choices are in fact equivalent as they differ by a trivial Wilson line shift.}
\be
e^{\frac{n_x}{M} D_x}\quad {\rm and} \quad e^{\frac{n_y}{M} D_y } \qquad \quad n_x, n_y = 0, \dots, M-1
\label{opschiral}
\ee
with these operators acting on the zero modes obtained by solving the internal Dirac or Laplace equations on $\T^2$ \cite{magnus}
\be
\label{eq:func}
\psi^{j,M}(z,\bar z)=
\left\{
\begin{array}{l}
\vspace*{.2cm}
 e^{i\pi M {z {\rm Im} z}/{{\rm Im} \tau}}
\, \vartheta\left [\begin{array}{c}
\frac{j}{M}\\
0\end{array}\right ](Mz,M \tau)
\qquad \text{if} \quad M>0\\
 e^{i\pi M {\bar{z} {\rm Im} \bar{z}}/{{\rm Im} \tau}}
\,\vartheta\left [\begin{array}{c}
\frac{j}{M}\\
0\end{array}\right ](M\bar{z},M \bar{\tau})
\qquad \text{if} \quad M<0
\end{array}\right.
\ee
$j=0,1,\dots,|M|-1$ running over independent zero mode solutions. One can check that 
\be
g_\cw^{n_x} =  e^{\frac{n_x}{M}D_x} \psi^{j,M}= e^{2\pi i \frac{n_x j}{M}}\psi^{j,M} \qquad g_\ct^{n_y} =  e^{\frac{n_y}{M}D_y} \psi^{j,M} = \psi^{j+n_y,M}
\label{glawdis}
\ee
so that if we consider the vector of wavefunctions
\be
{\Psi}\, =\, 
\left(
\begin{array}{c}
\psi^{0,M} \\  \vdots \\ \psi^{M-1,M}
\end{array}
\right)
\label{psivec}
\ee
we have that the group elements $g_\ct$ and $g_\cw$ act as (\ref{clock}) and (\ref{shift}) respectively,  generating the discrete Heisenberg group $H_M  \simeq ( \IZ_M \times \IZ_M) \rtimes \IZ_M$ as mentioned above.

If we now consider the full $\T^6 = (\T^2)_1 \times (\T^2)_2 \times (\T^2)_3$ magnetized D9-brane system, we obtain that the zero mode wavefunctions for the $D9_aD9_b$ sector are \cite{magnus}
\be
\psi_{ab}^{j_1, j_2, j_3}\, =\, \psi_{ab}^{j_1, - I_{ab}^1} (z_1, \bar{z}_1)  \cdot  \psi_{ab}^{j_2, - I_{ab}^2} (z_2, \bar{z}_2) \cdot  \psi_{ab}^{j_3, - I_{ab}^3} (z_3, \bar{z}_3)
\label{zmT6}
\ee
with $I_{ab}^i = m_b^i - m_a^i$ and $\psi^{j,M}$ as in (\ref{eq:func}). The number of zero modes in the $ab$ sector is given by $|I_{ab}^{\T^6}| = |I_{ab}^1||I_{ab}^2||I_{ab}^3|$ , as expected from the T-dual  intersecting D6-brane system, and their 4d chirality is again given by ${\rm sign} (I_{ab}^{\T^6})$. From our discussion on $\T^2$ it follows that each index $j_i$ transforms in the fundamental representation of $H_{I_{ab}^i}$. We then obtain again that the flavor symmetry group of this sector is given by ${\bf P}^{ab}_{\T^6} \, =\, H_{I_{ab}^1} \times  H_{I_{ab}^2} \times  H_{I_{ab}^3}$, as in (\ref{disT6P}). 

Let us now consider magnetized D9-branes in a $\T^6/\IZ_2 \times \IZ_2$ orbifold background,\footnote{More precisely, we consider the type IIB orbifold background mirror symmetric to our previous type IIA $\T^6/\IZ_2 \times \IZ_2'$ background. These two backgrounds are quite similar but not exactly the same, because upon three T-dualities the choice of discrete torsion of a $\IZ_2 \times \IZ_2$ orbifold is reversed. As a result, the fixed points of the type IIB $\IZ_2 \times \IZ_2$ orbifold considered in this section contain collapsed two and four-cycles instead of collapse three-cycles.} again with the $\IZ_2$  generators acting as (\ref{orbiaction}). Because of the presence of the orbifold fixed loci, the $U(1)^6$ translational symmetry of $\T^6$ is broken down  to the discrete subgroup $\IZ_2^6$, and this reduces the set of operators of the form (\ref{opschiral}) that are compatible with the symmetries of the background.

For our purposes it is instructive to again consider the toy example $\T^2/\IZ_2$ with $\IZ_2$ action generated by $z \mapsto -z$. As before, this background has four fixed points at $\{0, 1/2, \tau/2, (1+\tau)/2\}$ that are interchanged by the $\IZ_2 \times \IZ_2$ symmetry group generated by the discrete translations $z \mapsto z + 1/2$ and $z \mapsto z + \tau/2$. Let us now consider the magnetized $U(2)$ sector (\ref{fluxab}) in this background. The $\T^2/\IZ_2$ discrete isometry $z \mapsto z + 1/2$ is implemented by exp$(\oh D_x)$, while $z \mapsto z + \tau/2$ is implemented by exp$(\oh D_y)$. From the discussion above, we know that these operators correspond to symmetries of the magnetized system only if they belong to (\ref{opschiral}), or in other words if $M$ is even. We then find that a pair of magnetized D-branes respects the orbifold translational $\IZ_2 \times \IZ_2$ symmetry if and only if $I_{ab} = $ even, exactly as we found in the T-dual picture of intersecting D-branes.

As it is clear from (\ref{glawdis}), for $M$ even the group elements $g_\ct^{M/2}$ and $g_\cw^{M/2}$ will generate a discrete flavor symmetry group acting on the zero mode wavefunctions. Because the group action  is non-Abelian and in general it describes a discrete Heisenberg group, we can identify the flavor group with $H_2 \simeq ( \IZ_2 \times \IZ_2) \rtimes \IZ_2 \simeq D_4$. As we will see, the families of wavefunctions indeed arrange themselves in representations of the dihedral group $D_4$.

\subsection{Wavefunction representations in $\T^6/\IZ_2 \times \IZ_2$}

Let us consider in detail the wavefunctions for bifundamental fields in a magnetized D9-brane model in $\T^6/\IZ_2 \times \IZ_2$. In general the spectrum of zero modes will be similar to the case of $\T^6$, except that we need to project out those modes that are not invariant under the orbifold action. The procedure for finding the surviving chiral families works pretty much like in section (\ref{sub:orbichiral}) (see \cite{ako08} for a previous discussion). The chiral matter will have a specific SO(8) weight representation inherited from the 10d fields $(\Psi, A_M)$. In particular, for 4d massless fermions we also find the representation (\ref{rfermion}) with $s_i =  {\rm sign} (I_{ab}^i)$. The orbifold generators will act on the internal Lorentz indices of the chiral fermions as in (\ref{orbiferm}) and select either even or odd linear combinations of wavefunctions for each $(\T^2)_i$. We will again have wavefunctions of the form (\ref{zmT6}) but with (\ref{eq:func}) replaced by
\be
\label{wfparity}
\psi_{\text{even}}^j \propto \psi^{j,M}+\psi^{M-j,M}\qquad {\rm or} \qquad
\psi_{\text{odd}}^j \propto \psi^{j,M}-\psi^{M-j,M}
\ee
depending on each case and $(\T^2)_i$. The family indices $j_i$ will run over $I_e^i$ or $I_o^i$ different values, cf.(\ref{evenodd}), and so the total number of chiral families will again be given by (\ref{topint}). 

In the following we will analyze the different representations of the flavor group under which the chiral families transform. Since each index in $\psi_{ab}^{j_1j_2j_3} = \psi_{ab}^{j_1}\psi_{ab}^{j_2}\psi_{ab}^{j_3}$ transforms independently we can treat each wavefunction factor $\psi_{ab}^{j_i}$ separately, which is equivalent to consider the representations of even and odd wavefunctions in our toy example $\T^2/\IZ_2$. As we now show, the decomposition into irreducible representations of $D_4$ depends on the value of the $\T^2$ magnetic flux $M$, which we assume an even number

\subsubsection*{M=2}

In this case we have that
\be
\Psi_{\rm even}^{M=2} \, =\, 
\left(
\begin{array}{c}
\psi^{0,2} \\  \psi^{1,2}
\end{array}
\right)
\label{psivec2}
\ee
while all odd wavefunctions are projected out. The group elements $g_\cw$ and $g_\ct$ generate the $2\times 2$ irreducible representation of $D_4$, given by
\bea\label{R2}
M_1=\left ( \begin{array}{cc}
1&0\\
0&1\end{array}\right ) ,\quad M_2=\left ( \begin{array}{cc}
0&$-$1\\
1&0\end{array}\right ), \quad M_3=\left ( \begin{array}{cc}
$-$1&0\\
0&$-$1\end{array}\right ), \quad M_4=\left ( \begin{array}{cc}
0&1\\
$-$1&0\end{array}\right )\\\nonumber
M_5=\left ( \begin{array}{cc}
1&0\\
0&$-$1\end{array}\right ) ,\quad M_6=\left ( \begin{array}{cc}
0&1\\
1&0\end{array}\right ), \quad M_7=\left ( \begin{array}{cc}
$-$1&0\\
0&1\end{array}\right ), \quad M_8=\left ( \begin{array}{cc}
0&$-$1\\
$-$1&0\end{array}\right )
\eea
More precisely, we have that $g_\cw = M_5$, $g_\ct = M_6$ and that all other elements are generated by multiplication of these two. In the following we will refer to this 2-dimensional, faithful irrep of $D_4$ as $\mathbf R_2$.

\subsubsection*{M=4}

For $M=4$ we have that 
\be
\Psi_{\rm even}^{M=4} \, =\, 
\left(
\begin{array}{c}
\psi^{0,4} \\ \frac{1}{\sqrt{2}} (\psi^{1,4} + \psi^{3,4}) \\ \psi^{2,4}
\end{array}
\right)
\qquad {\rm and} \qquad 
\Psi_{\rm odd}^{M=4} \, =\, \frac{1}{\sqrt{2}} (\psi^{1,4} - \psi^{3,4})
\label{psivec4}
\ee
and the generators of the group action are given by $T_C = g_\cw^2$ and $T_S = g_\ct^2$. On the three dimensional vector of even wavefunctions these elements read
\be
T_C\, =\,
\left ( \begin{array}{ccc}
1&0 & 0\\
0& -1 & 0 \\
0 & 0 & 1
\end{array}\right )
\qquad {\rm and} \qquad
T_S\, =\,
\left ( \begin{array}{ccc}
0 & 0 & 1\\
0& 1 & 0 \\
1 & 0 & 0
\end{array}\right )
\ee
which is not an irreducible representation: we can consider a new basis of even wavefunctions
\be
\tilde{\Psi}_{\rm even}^{M=4} \, =\,  \frac{1}{\sqrt{2}}
\left(
\begin{array}{c}
\psi^{0,4} + \psi^{2,4}  \\ \psi^{1,4} + \psi^{3,4} \\ \psi^{0,4} - \psi^{2,4}
\end{array}
\right)
\label{psivec4b}
\ee
in which $T_C = {\rm diag\, } (1,-1,1)$ and $T_S = {\rm diag\, } (1,1,-1)$. This representation of $D_4$ is thus equivalent to
\be
\mathbf (+,+) \oplus (-,+) \oplus (+,-)
\ee
where $(-,+)$ is the one-dimensional representation of $D_4$ such that $T_C = 1$ and $T_S = -1$. One can also see that $D_4$ acts on $\Psi_{\rm odd}$ as a one-dimensional representation such that $T_C = T_S = -1$, or using the above notation as
\be
(-,-) \,\simeq \,{\rm det\, } \mathbf R_2
\ee

\subsubsection*{M=2k}

For general $M=2k$, $k \in \IZ$ it is convenient to define the even wavefunctions as
\bea\nonumber
\xi_e^{0,k}&=&\psi^{0,2k}\\
\xi_e^{j,k}&=&\frac{1}{\sqrt 2} (\psi^{j,2k}+\psi^{2k-j,2k}) \qquad\text{for}\quad j=1,2,\dots, k-1\\\nonumber
\xi_e^{k,k}&=&\psi^{k,2k} 
\eea
One can check that the flavor group generators act on them as follows
\bea\nonumber
T_C \,\xi_e^{j,k}&=&(-1)^j\xi_e^{j,k}\\\nonumber
T_S\,\xi_e^{j,k}&=&\xi_e^{k-j,k}
\eea
where we have defined $T_C = g_\cw^k$ and $T_S = g_\ct^k$. In matrix terms one obtains the following $(k+1)$-dimensional representation.
\be
\Psi_{\rm even}^{M=2k} \equiv\left ( \begin{array}{c}
\xi_e^{0,k}\\
\vdots\\
\xi_e^{k,k} \end{array}\right ), \ T_C =\left (\begin{array}{cccc}
1&0&\dots&0\\
0&-1&\dots&0\\
\vdots&\vdots&&\vdots\\
0&0&\dots&(-1)^k
\end{array}\right ),\ T_S=\left (\begin{array}{cccc}
0&\dots&0&1\\
0&\dots&1&0\\
\vdots&&\vdots&\vdots\\
1&\dots&0&0
\end{array}\right )
\ee

One can see that for $k= M/2$ even the two generators $T_C$ and $T_S$ commute, so they can be simultaneously diagonalized. This diagonalization corresponds to decompose the original $(k+1)$-dimensional representation into a sum of one-dimensional ones. In this case one gets the following decomposition:
\be
\begin{array}{lcl}
k=4s& \longrightarrow& \mathbf{k+1}=\oplus_{i=1}^{s+1} (+,+)_i \oplus_{j=1}^s (+,-)_j \oplus_{k=1}^{s} (-,+)_k \oplus_{l=1}^s (-,-)_l\\
k=4s+2& \longrightarrow& \mathbf{k+1}= \oplus_{i=1}^{s+1} (+,+)_i \oplus_{j=1}^{s+1} (+,-)_j \oplus_{k=1}^{s+1} (-,+)_k \oplus_{l=1}^s (-,-)_l
\end{array}
\ee
where $(\eps_1,\eps_2)$ stands for the one-dimensional representation in which $T_C = \eps_1$ and $T_S = \eps_2$. 

Let us now examine the case in which $k=2s+1$ is odd. The matrices $T_C$ and $T_S$ are $(2s+2)$-dimensional and they do not commute in this case, but they are still diagonalizable by blocks. Every block is the same and identical to the $2 \times 2$ matrices in (\ref{R2}), which means that we can write $T_C =  g_\cw^k$ and $T_S = g_\ct^k$ as $\oplus^{s+1} M_5$ and $\oplus^{s+1} M_6$, respectively. In other words the flavor group action can be expressed as
\be
\oplus_{i=1}^{s+1} (\mathbf R_2)_i
\ee

For the case of odd wavefunctions we can take the following definitions
\be
\xi_o^{j,k}\equiv \frac{1}{\sqrt{2}}(\psi^{j,2k}-\psi^{2k-j,2k}) \qquad\text{for}\quad j=1,2,\dots,k-1.
\ee
and check that in this case the group generators act as
\be
\Psi_{\rm odd}^{M=2k} \equiv\left ( \begin{array}{c}
\xi_o^{1,k}\\
\vdots\\
\xi_o^{k-1,k} \end{array}\right ), \ T_C =\left (\begin{array}{cccc}
-1&0&\dots&0\\
0&+1&\dots&0\\
\vdots&\vdots&&\vdots\\
0&0&\dots&(-1)^{k-1}
\end{array}\right ),\ T_S=\left (\begin{array}{cccc}
0&\dots&0&-1\\
0&\dots&-1&0\\
\vdots&&\vdots&\vdots\\
-1&\dots&0&0
\end{array}\right )
\ee

Where again $T_C = g_\cw^k$ and $T_S = g_\ct^k$. For $k$ even these two matrices commute and one has an Abelian $(k-1)$-dimensional representation decomposable as
\be
\begin{array}{lcl}
k=4s& \longrightarrow& \mathbf{k-1}=\oplus_{i=1}^{s-1} (+,+)_i \oplus_{j=1}^s (+,-)_j \oplus_{k=1}^{s} (-,+)_k \oplus_{l=1}^s (-,-)_l\\
k=4s+2& \longrightarrow& \mathbf{k-1}= \oplus_{i=1}^{s} (+,+)_i \oplus_{j=1}^{s} (+,-)_j \oplus_{k=1}^{s} (-,+)_k \oplus_{l=1}^{s+1} (-,-)_l
\end{array}
\ee

\begin{table}[htb] 
\renewcommand{\arraystretch}{1.2}
 \begin{center}
    \begin{tabular}{ | c || c | c |}
    \hline
 $M$ & $\Psi_{\rm even}$\quad  \footnotesize{${\rm dim} = M/2 + 1$}& $\Psi_{\rm odd}$  \quad \footnotesize{${\rm dim} = M/2 - 1$} \\ \hline\hline
 {\footnotesize $4s+2$} &  $ \stackrel{s+1}{\oplus}  \mathbf R_2$ & $ \stackrel{s}{\oplus}  \mathbf R_2$\\ \hline
{\footnotesize $8s+4$} & $\stackrel{s+1}{\oplus} (+,+) \stackrel{s+1}{\oplus} (+,-) \stackrel{s+1}{\oplus} (-,+) \stackrel{s}{\oplus} (-,-)$ & $\stackrel{s}{\oplus} (+,+) \stackrel{s}{\oplus} (+,-) \stackrel{s}{\oplus} (-,+) \stackrel{s+1}{\oplus} (-,-)$   \\ \hline
    {\footnotesize $8s+8$} & $\stackrel{s+2}{\oplus} (+,+) \stackrel{s+1}{\oplus} (+,-) \stackrel{s+1}{\oplus} (-,+) \stackrel{s+1}{\oplus} (-,-)$  & $\stackrel{s}{\oplus} (+,+) \stackrel{s+1}{\oplus} (+,-) \stackrel{s+1}{\oplus} (-,+) \stackrel{s+1}{\oplus} (-,-)$   \\ \hline
    \end{tabular}
\end{center}
\caption{Different family representations depending on the value of the magnetization $M \in 2\IZ$ for even and odd wavefunctions. Here $\mathbf R_2$ stands for the 2-dimensional irreducible representation of the dihedral group $D_4$, as in  (\ref{R2}).}
\label{tab:repsM}
\end{table}

Finally, for $k=2s+1$ odd the group generators do not commute, but just like in the case of even wavefunctions the $2s$-dimensional representation is reducible into $s$ copies of the two-dimensional representation $\mathbf R_{2}$. We then have that the flavor group acts as
\be
\oplus_{i=1}^{s} (\mathbf R_2)_i
\ee
All these results have been summarized in table \ref{tab:repsM}. In the next section we will apply them to see how families of quarks and leptons transform in specific semi-realistic models.
%


\section{Examples}
\label{s:example}

In this section we illustrate our general analysis via a couple of semi-realistic examples. More precisely, we will consider two intersecting D6-brane models on the $\IZ_2 \times \IZ_2'$ orbifold with a Pati-Salam gauge group. The first example is a four generation model already constructed in \cite{bcms05}, with a $D_4 \times D_4 \times D_4$ symmetry group constraining its Yukawa couplings. The second example is a new, three generation model with a $D_4$ symmetry group. 

One important ingredient of these models is the presence of orientifold planes, that allow to construct consistent and stable D-brane configurations. While the presence of O-planes does not change the discrete symmetries of a $\IZ_2 \times \IZ_2'$ orbifold background, it does affect the D-brane content of a model and the associated 4d chiral spectrum. Hence, before presenting our examples we briefly review the effect of adding an orientifold projection to the $\IZ_2 \times \IZ_2'$ orbifold. 

\subsection{Orientifolding}

In general, in order to build consistent, stable and 4d Poincar\'e invariant models based on intersecting or magnetized D-branes in Calabi-Yau compactifications we need to include the presence of negative tension objects that cancel the positive tension of the D-branes. The simplest way to do so is to include the presence of the non-dynamical, negative tension objects known as orientifold planes. In the case of type IIA string theory compactified on $(\T^2)_1 \times (\T^2)_2 \times (\T^2)_3$ this is achieved by modding out the theory by $\Om\CR$, where $\Om$ stands for the worldsheet parity operator and $\CR$ for the anti-holomorphic involution $\CR\, :\, z_i \mapsto \bar{z}_i$. For this the D-brane configuration has to be invariant under the action of $\Om\CR$, and so for each D6-brane wrapping the three-cycle (\ref{torcycle}) there must be another D6-brane wrapping $\Pi_{\a'} = \CR \Pi_\a$. If as in \cite{bgkl00} we consider that each $(\T^2)_i$ has a rectangular geometry, a $U(N_a)$ gauge group will arise from wrapping $N_a$ D6-branes on $\Pi_a$ and also on $\Pi_{a'}$, where
\be
\begin{array}{lcccc}
\Pi_{a} & : & (n_a^1, m_a^1) &  (n_a^2, m_a^2)  & (n_a^3, m_a^3) \\
\Pi_{a'} & : & (n_a^1, -m_a^1) & (n_a^2, -m_a^2) & (n_a^3, -m_a^3) 
\end{array}
\label{D6aa'}
\ee
In order to obtain a gauge group $U(N_a) \times U(N_b)$ we also need to place $N_b$ D6-branes on $\Pi_b$ and $\Pi_{b'}$. The spectrum of 4d left-handed chiral fermions in bifundamental representations is then given by \cite{bgkl00}
\be
I_{ab}^{\T^6} (N_a, \bar{N}_b) \quad + \quad I_{ab'}^{\T^6} (N_a, N_b)
\label{intab'}
\ee
where $I_{ab'}^{\T^6}  =  I_{ab'}^1I_{ab'}^2I_{ab'}^3 = \prod_{i=1}^3 (n_a^i m_b^i + n_b^im_a^i)$. In addition there are 4d chiral fermions arising from the intersection of $\Pi_a$ with its orientifold image $\Pi_{a'}$, that transform in the symmetric and antisymmetric representation of $U(N_a)$, namely we have
\be
\Yasymm_a\, \oh (I_{aa'}^{\T^6} - 8 m_a^1m_a^2m_a^3) \quad + \quad \Ysymm_a \,\oh (I_{aa'}^{\T^6} + 8 m_a^1m_a^2m_a^3)
\label{intaa'}
\ee

The same orientifold projection can be performed for the toroidal orbifold $\T^6/\IZ_2 \times \IZ_2'$, also by modding out the theory by $\Om\CR$. Again, a rigid D6-brane wrapping $\Pi_a$ will have an orientifold image wrapping $\Pi_{a'}$. It is easy to see that $\Pi_a$ and $\Pi_{a'}$ will go through the same fixed points on each $(\T^2)_i$, and so adding D6-brane orientifold images will not break the discrete flavor symmetry of the model any further.  In order to obtain the chiral spectrum in this background one must consider (\ref{intab'}) and (\ref{intaa'}) and project out all the chiral modes that are not invariant under the orbifold action. Following our discussion of section \ref{sub:orbichiral}, it is easy to see that (\ref{intab'}) is replaced by
\be
I_{ab} (N_a, \bar{N}_b) \quad + \quad I_{ab'} (N_a, N_b)
\label{topintab'}
\ee
where $I_{ab}$ is given by (\ref{topint}) and similarly for $I_{ab'}$ with the replacement $b\raw b'$. The orbifold projection of (\ref{intaa'}) is less straightforward but one can check that it amounts to
\be
\Yasymm_a\, \oh (I_{aa'} +  4\, I_{aO6}^{\T^6}) \quad + \quad \Ysymm_a \,\oh (I_{aa'} -  4\, I_{aO6}^{\T^6})
\label{topintaa'}
\ee
where for computing $I_{aa'}$ we use again the expression (\ref{topint}), but with the wrapping numbers of $\Pi_{a'}$ instead of $\Pi_b$. On the other hand, $I_{aO6}^{\T^6}$ is the $\T^6$ intersection number (\ref{topintT6}) between $\Pi_a$ and the three-cycle $\Pi_{O6}$, with
\be
[\Pi_{O6}]\, =\, - 2\left(  [a^1]\cdot[a^2]\cdot[a^3] +  [a^1]\cdot[b^2]\cdot[b^3] +  [b^1]\cdot[a^2]\cdot[b^3] +  [b^1]\cdot[b^2]\cdot[a^3] \right)
\label{3cycleO6}
\ee

The three-cycle (\ref{3cycleO6}) has a geometrical interpretation, namely that the orientifold projection $\Om\CR$ introduces a set of O6-planes that are located at the fixed point loci of $\Om\CR$, $\Om\CR\Theta$, $\Om\CR\Theta'$ and $\Om\CR\Theta\Theta'$, and adding up the homology classes of all these three-cycles we can associate a total homology class $[\Pi_{O6}]$ for the O6-plane. If each $(\T^2)_i$ has a rectangular geometry such homology class is given by (\ref{3cycleO6}). We refer the reader to \cite{bcms05} for other cases in which some $(\T^2)_i$ is not rectangular, and for a generalization of eqs.(\ref{topintab'}), (\ref{topintaa'}) to these cases.\footnote{Our conventions are such that a positive intersection number $I_{ab}$ signals a net amount of $|I_{ab}|$ 4d left-handed chiral fermions in the representation $(N_a, \bar{N}_b)$, while a negative intersection signals $|I_{ab}|$ fermions in the same representation but with opposite chirality. In \cite{bcms05} this chirality convention is reversed.}

The importance of introducing O6-planes is that they allow to construct consistent and supersymmetric D6-brane models \cite{csu01}. In general, a D6-brane model will be consistent if and only if the RR-tadpole condition 
\be
\sum_\a N_\a ([\Pi_\a^F] + [\Pi^{F}_{\a^\prime}]) \, =\,  4 [\Pi_{O6}] 
\label{RRtadpoles}
\ee
is satisfied. Here the index $\a$ runs over each of the D6-branes of the model, $\Pi_\a^F$ stands for the fractional three-cycles described in appendix \ref{ap:review} and $\Pi^{F}_{\a^\prime}$ is the image of $\Pi_\a^F$ under $\CR$. As discussed in appendix \ref{ap:review} we can describe a D6-brane on  $\Pi_\a^F$ in terms of $\T^6$ wrapping numbers $(n_\a^i, m_\a^i)$. One can then see that the condition for a D6-brane model to preserve the $\CN=1$ supersymmetry of the $\T^6/\IZ_2 \times \IZ_2' \times \Om\CR$ background is \cite{bdl96}
\be
\th_\a^1 + \th_\a^2 + \th_\a^3\, =\, 0 \ {\rm mod\ } 2\pi, \quad \forall\, \a 
\label{angle}
\ee
with $\th_\a^i \, =\, {\rm tan}^{-1} \frac{m_\a^i R_{y_i}}{n_\a^i R_{x_i}}$. As shown in \cite{bcms05}, both conditions (\ref{3cycleO6}) and (\ref{angle}) are equivalent to simple expressions in terms of the $\T^6$ wrapping numbers $(n_\a^i, m_\a^i)$ and can be satisfied simultaneously. In the next subsection we will consider a set of D6-branes which are a Pati-Salam subsector of a D6-brane model built in \cite{bcms05} satisfying both conditions.

\subsection{A global Pati-Salam four-generation model}

As an example of D-brane model with non-trivial flavor symmetry group let us consider the intersecting D6-brane model in table 8 of \cite{bcms05}, which is based on the orientifold background $\T^6/\IZ_2 \times \IZ_2' \times \Om \CR$. In particular, we will consider the subsector given by the D6-branes $a_1$, $a_2$ and $a_3$ in that model, whose wrapping numbers we display in table \ref{tab:aaa}. 
\begin{table}[htb]
\renewcommand{\arraystretch}{1}
\begin{center}
	\begin{tabular}{| c | c | c | c |}
	\hline
	$N_\a$ & $(n_\a^1,m_\a^1)$ & $(n_\a^2,m_\a^2)$ & $(n_\a^3,m_\a^3)$ \\ \hline \hline
	$N_{a_1}=4$ & $(1,0)$ & $(0,1)$ & $(0,-1)$ \\\hline
	$N_{a_2} =2$ & $(1,0)$ & $(2,1)$ & $(4,-1)$ \\\hline
	$N_{a_3}=2$ & $(-3,2)$ & $(-2,1)$ & $(-4,1)$ \\ \hline
	\end{tabular}
	\end{center}
\caption{Wrapping numbers for the four-generation Pati-Salam model of \cite{bcms05}.}
\label{tab:aaa}
\end{table}

The gauge group that arises from this set of D-branes is given by $U(4) \times U(2) \times U(2)$, and as shown in \cite{bcms05} the chiral spectrum contains four families of left-handed chiral fermions in the representations $\mathbf{(4,2,1)+(\bar{4},1,2)}$. We then have a four-generation Pati-Salam model, and because the supersymmetry conditions (\ref{angle}) amount to impose
\be
\begin{array}{c}
2U^2\, =\, U^3\\
{\rm tan}^{-1} \left(\frac{2U^1}{3}\right) +{\rm tan}^{-1} \left(\frac{U^2}{2}\right)  + {\rm tan}^{-1} \left(\frac{U^3}{4}\right) \, =\, \pi
\end{array}
\ee
with $U^i = R_{y_i}/R_{x_i}$, we can find a continuum of supersymmetric solutions. The matter spectrum then contains $4\mathbf{(4,2,1)}+4\mathbf{(\bar{4},1,2)}$ $\CN=1$ left-handed chiral multiplets. 

\begin{table}[htb] 
\renewcommand{\arraystretch}{1}
 \begin{center}
    \begin{tabular}{ | c || c || c | c | c || c | c |}
    \hline
    Sector & ${U(4)\times U(2)_L\times U(2)_R}$ &$I_{\a\b}^1$ & $I_{\a\b}^2$ & $I_{\a\b}^3$ &Projection  & $I_{\a\b}$ \\ \hline\hline
      $a_1a_2$ &$({\mathbf 4},\bar{\mathbf 2},\mathbf 1)$& 0& -2& 4&$- I^2_e\,I_o^3$  & 2 \\ \hline
      $a_1a_2'$ &$({\mathbf 4},{\mathbf 2},\mathbf 1)$& 0 & -2 & 4 &$- I^2_e\,I_o^3$ & 2   \\ \hline \hline
      $a_1a_3$ &$(\mathbf 4,\mathbf 1,\bar{\mathbf 2})$& 2  & 2 & -4 &$I^1_e\,I^2_e\,I_o^3$ & -4   \\ \hline
      $a_1a_3'$ &$(\mathbf 4,\mathbf 1,\mathbf 2)$& -2 & 2 & -4 & $-$ & 0 \\ \hline \hline
      $a_2a_3$ &$(\mathbf 1,\mathbf 2,\bar{\mathbf 2})$& 2  & 4 & 0 &$I^1_e\,I^2_e$ & 6  \\ \hline
      $a_2a_3'$ &$(\mathbf 1,\mathbf 2,\mathbf 2)$& -2 & 0 & -8 & -$I^1_e\,I_e^3$ & -10  \\ \hline \hline
      $a_2a_2'$ &$(\mathbf 1,\mathbf 1_{+2},\mathbf 1)$& 0 & -4 & 8 &$I^2_o\,I^3_e- I^2_e\,I_o^3$ & -5+9  \\ \hline
      $a_3a_3'$ &$(\mathbf 1,\mathbf 1,\mathbf 1_{+2})\, \&  (\mathbf 1,\mathbf 1,\mathbf 3)$& 12 & 4 & 8 &$I^1_e\,I^2_e\,I_e^3+I^1_o\,I^2_o\,I_o^3$ & 105+15  \\ \hline 
      \end{tabular}
\end{center}
\caption{Bulk intersection numbers together and the wavefunctions surviving the orbifold action in the four-generation Pati-Salam model. The last column shows the total intersection number. A positive intersection number indicates a left-handed $\CN=1$ chiral multiplet, and a negative one a right-handed chiral multiplet.}
\label{tab:sector4}
\end{table}

\subsubsection*{Flavor group and representations}

Let us analyze this Pati-Salam model in light of the results of section \ref{s:review}.
Table \ref{tab:sector4} shows the toroidal intersection numbers $I^i_{\a\b}$ for each of the relevant sectors of this model. Notice that all these numbers are even, and so in (\ref{disZ2P2}) one has that $d_1=d_2=d_3=2$. That is, the flavor symmetry group of this sector is given by
\be
{\bf P}^{a_1a_2a_3}_{\T^6/\IZ_2 \times \IZ_2'} \, =\, D_4^{(1)} \times  D_4^{(2)} \times D_4^{(3)}
\label{disZ2Ex4}
\ee
where the factor $D_4^{(i)}$ arises from the symmetries on the two-torus $(\T^2)_i$. While (\ref{disZ2Ex4}) corresponds to a symmetry of the Pati-Salam sector, it does not need to be respected by the whole D-brane model. Indeed, in order to satisfy the consistency conditions (\ref{RRtadpoles}) we will need to add extra sets of D6-branes to those of table \ref{tab:sector4}, and in order for (\ref{disZ2Ex4}) to be an exact symmetry of the model all the intersection numbers $I^i_{\a\b}$ involving these extra D6-branes  also need to be even. In general one would not expect this to be the case and, indeed, by looking at the completion of this model given by table 8 of \cite{bcms05} one realizes that there is always some D6-brane $\b$ of this extra set such that $I^i_{a_j\b}=$ odd for any given $i$. The flavor symmetry group (\ref{disZ2Ex4}) is then broken by the presence of the other D-branes of this model, and can only be thought as an approximate symmetry of the Pati-Salam sector of table \ref{tab:sector4}. Nevertheless, even if not exact this symmetry will constrain the Yukawa couplings of this model at tree-level and, because of $\CN=1$ supersymmetry, at all orders in perturbation theory. It is then useful to analyze under which representation of the discrete flavor group (\ref{disZ2Ex4}) transform each of the chiral modes of table \ref{tab:sector4}. 

Let us for instance consider the sector $a_1a_2$ of this model, which contains 2 copies of left-handed multiplets in the representation $(\mathbf4,\bar{\mathbf 2},\mathbf 1)$. In order to see how these two copies arise we must compute the combination of even or odd points on each two-torus that survive the orbifold projection. Following our discussion of section \ref{s:review} and due to the particular signs of $I_{a_1a_2}^i$ for $i=1,2,3$ one is instructed to keep the wavefunctions of the form $\psi_{\rm even}^{j_2}\psi_{\rm odd}^{j_3}$ or $\psi_{\rm odd}^{j_2}\psi_{\rm even}^{j_3}$, counted with the appropriate chirality. More precisely, the chiral index in this sector is given by $I_o^2I_e^3-I_e^2I_o^3=0-(-2)=2$,  in agreement with \cite{bcms05}. In fact, because $I_{a_1a_2}^2=-2$ there are no odd wavefunctions in $(\T^2)_2$ and so the wavefunctions that correspond to this sector are of the form
\be
\psi_{a_1a_2}^{j_2}\, =\, \psi_{\rm even}^{j_2}\cdot \psi_{\rm odd}^{j_3} \quad \quad j_2 =0,1,\quad j_3=0
\label{zmEx4}
\ee
By looking at table \ref{tab:reps} one can see how these chiral modes transform under the flavor group ${\bf P}^{a_1a_2a_3}$. On the one hand the index $j_2$ transforms in the 2-dimensional representation $\mathbf R_2$ of $D_4^{(2)}$, and on the other hand the index $j_3$ only takes one value and transforms as $(-,-)$ under the flavor subgroup $D_4^{(3)}$. 

\begin{figure}[htb]
  \centering
  \def\svgwidth{460pt}
  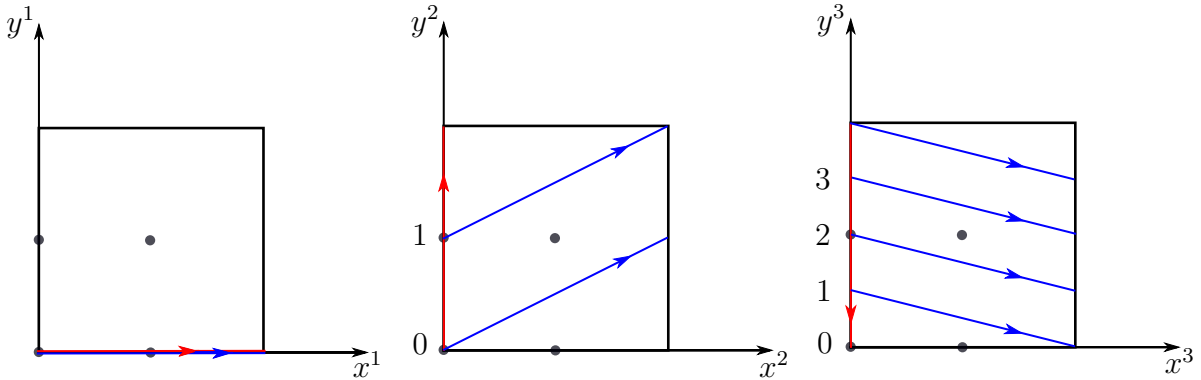
  \caption{Branes $a_1$ (red) and $a_2$ (blue) with labels for the different intersection points.}
  \label{fig:frozen}
\end{figure}

Geometrically, one can understand this result by drawing both D6-branes and labelling their intersection points as $p_2^{j}$ with $j=0,1$ in $(\T^2)_2$ and $p_3^{k}$ with $k=0,1,2,3$ in $(\T^2)_3$, see Figure \ref{fig:frozen}. Clearly, the two points in the second torus are even under the orbifold action and there are no odd points. In the third torus we find three even points, namely, $p_3=\{0,1+3,2\}$ and an odd one given by $p_3=\{1-3\}$. Since the orbifold action selects the points whose parity are (odd,even) or (even,odd), we have that the surviving points are $(p_2,p_3)=(\{0\},\{1-3\})$ and $(p_2,p_3)=(\{1\},\{1-3\})$ which correspond to the two different chiral modes in (\ref{zmEx4}). One can now see how the translation $z_2 \mapsto z_2 + \tau/2$ interchanges these two points, while they pick up a minus sign under the translation $z_3 \mapsto z_3 + \tau/2$. Adding up the action of the discrete B-field transformation (see footnote \ref{footshift}) we indeed recover that these two zero modes transform as $\mathbf R_2 \otimes (-,-)$ under $D_4^{(2)} \times D_4^{(3)}$. Finally, it is easy to see that the D6-brane intersections, which are the line $\{y_1 = 0\}$ in $(\T^2)_1$, are invariant under the translation $z_1 \mapsto z_1 +1/2$ and in general by the full action of $D_4^{(1)}$. The final result has been summarized in table \ref{tab:disc}, together with the representations for the other sectors of the form $a_ia_j$ and $a_ia_{j'}$ with $i\neq j$, that can be treated similarly.
\begin{table}[htb] 
\renewcommand{\arraystretch}{1}
 \begin{center}
    \begin{tabular}{ | c | c || c | c | c |}
    \hline
        Sector & Field & $D_4^{(1)}$  & $D_4^{(2)}$ & $D_4^{(3)}$  \\ \hline\hline
      $a_1a_2$ & $F_L=({\mathbf 4},\bar{\mathbf 2},\mathbf 1)$ & $\mathbf 1$ & $\mathbf R_2$ & $(-,-)$ \\ \hline
      $a_1a_2'$ & $F_L'=({\mathbf 4},{\mathbf 2},\mathbf 1)$ & $\mathbf 1$ &  $(-,-)$ & $\mathbf R_2$ \\ \hline 
      $a_1a_3$ & $F_R=(\bar{\mathbf 4},\mathbf 1,{\mathbf 2})$ & $\mathbf R_2$ & $\mathbf R_2$ & $(-,-)$  \\ \hline
      $a_2a_3$ & $H=(\mathbf 1,\mathbf 2,\bar{\mathbf 2})$ & $\mathbf R_2$ & $\mathbf 1\oplus(+,-)\oplus(-,+)$ & $\mathbf 1$ \\ \hline
      $a_2a_3'$ & $H'=(\mathbf 1,\bar{\mathbf 2},\bar{\mathbf 2})$ & $\mathbf R_2$ & $\mathbf 1$ & $\mathbf 1^2\oplus(+,-)\oplus(-,+)\oplus(-,-)$ \\ \hline 
      \end{tabular}
\end{center}
\caption{Representations of the Pati-Salam fields under the flavor symmetry group.}
\label{tab:disc}
\end{table}

\subsubsection*{Yukawa couplings}

Given the above representations under the flavor symmetry group one can now consider the Yukawa couplings
\be\label{eq:yuk}
\begin{array}{rcl}
Y:(a_1a_2)\otimes(a_1a_3)\otimes(a_2a_3)&\longrightarrow& ({\mathbf 4},\bar{\mathbf 2},\mathbf 1)\otimes (\bar{\mathbf 4},\mathbf 1,{\mathbf 2})\otimes (\mathbf 1, {\mathbf 2}, \bar{\mathbf 2})\\
Y':(a'_1a_2)\otimes(a_1a_3)\otimes(a_2'a_3)&\longrightarrow& ({\mathbf 4},{\mathbf 2},\mathbf 1)\otimes (\bar{\mathbf 4},\mathbf 1,{\mathbf 2})\otimes (\mathbf 1,\bar{\mathbf 2},\bar{\mathbf 2})
\end{array}
\ee
which are allowed by gauge invariance.\footnote{This includes those Abelian discrete gauge symmetries that remain after the $U(1)$ factors of the gauge group are broken by a St\"uckelberg mechanism \cite{bisu11}.} It however happens that several of these couplings are not allowed by the discrete flavor symmetry (\ref{disZ2Ex4}), as we will now see.

Let us first consider the coupling $Y$ in (\ref{eq:yuk}). In principle, $Y$ has $2\times 4\times 6$ independent components since there are $6$ different Higgses that appear in this set of Yukawa couplings. Nevertheless, in general the discrete symmetries in each torus will reduce the number of independent Yukawas. On the one hand, invariance under $D_4^{(1)}$ forces us to choose the singlet in $\mathbf 1\otimes\mathbf R_2\otimes\mathbf R_2=\mathbf 1\oplus(+,-)\oplus(-,+)\oplus(-,-)$ which reduces by a factor 4 the number of independent Yukawas. On the other hand, under $D_4^{(2)}$ the coupling $Y$ behaves as follows
\bea\nonumber
\mathbf R_2\otimes\mathbf R_2\otimes (\mathbf 1\oplus (+,-)\oplus(-,+))&=&(\mathbf 1\oplus(+,-)\oplus(-,+)\oplus(-.-))\otimes (\mathbf 1\oplus (+,-)\oplus(-,+))\\\nonumber
&=&\mathbf 1^3\oplus(+,-)^3\oplus(-,+)^3\oplus(-,-)^3
\eea
which reduces by another factor of ${4}$ the number of independent Yukawas. Finally, since $D_4^{(3)}$ does not impose further constraints we conclude that there are only $\frac{2\times 4\times 6}{4\times 4}= 3$ independent components in $Y$. In other words, at tree-level there will only be three independent Yukawas within this sector. More precisely one obtains the following Yukawa couplings $Y_{ijk} F_{L, i} F_{R, j} H_k$ where 
\be
Y_{ijk} H_k\, =\, 
\left ( \begin{array}{cccc}
a H_0+ c H_2&bH_1&a H_3+ c H_5& bH_4\\
bH_1 &aH_0 - cH_2&bH_4&a H_3- c H_5\end{array}\right ) 
\label{Yukmat}
\ee
where the row index $i$ runs over the two families of left-handed multiplets $F_L$ in the $a_1a_2$ sector, while the index $j$ runs over the four families of right-handed multiplets $F_R$. For concreteness we have displayed the definition of these multiplets in terms of D-brane intersections in table \ref{tab:wave}.
\begin{table}[htb] 
\renewcommand{\arraystretch}{1}
 \begin{center}
    \begin{tabular}{| c | c | c |}
    \hline
       $F_{L, i}$  & $F_{R, j}$ & $H_k$  \\ \hline\hline
       $(\psi^0)_2 \cdot (\psi^1 - \psi^3)_3$ & $(\psi^0)_1 \cdot (\psi^0)_2 \cdot (\psi^1 - \psi^3)_3$ & $(\psi^0)_1 \cdot (\psi^0+\psi^2)_2$ \\
        $(\psi^1)_2 \cdot (\psi^1 - \psi^3)_3$  &  $(\psi^0)_1 \cdot (\psi^1)_2 \cdot (\psi^1 - \psi^3)_3$ & $(\psi^0)_1 \cdot (\psi^0-\psi^2)_2$ \\ 
          &  $(\psi^1)_1 \cdot (\psi^0)_2 \cdot (\psi^1 - \psi^3)_3$ & $(\psi^0)_1 \cdot (\psi^1+\psi^3)_2$ \\
          &  $(\psi^1)_1 \cdot (\psi^1)_2 \cdot (\psi^1 - \psi^3)_3$ & $(\psi^1)_1 \cdot (\psi^0+\psi^2)_2$ \\
          & & $(\psi^1)_1 \cdot (\psi^0-\psi^2)_2$ \\
          & & $(\psi^1)_1 \cdot (\psi^1+\psi^3)_2$ \\
       \hline
      \end{tabular}
\end{center}
\caption{Wavefunctions of the fields in the Yukawa couplings (\ref{Yukmat}). Here $(\psi^j)_i$ stands for a delta-function localized at the $j^{th}$ intersection of the D6-branes $a_1$ and $a_2$ in $(\T^2)_i$.}
\label{tab:wave}
\end{table}

Considering now the Yukawa couplings $Y'$ in (\ref{eq:yuk}), one finds that the effect of the discrete flavor symmetry is even more dramatic since there is no combination which is invariant under the factor $D_4^{(2)}$. As a result these Yukawa couplings will vanish   and (\ref{Yukmat}) will be the only set of Yukawas at the perturbative level. Hence, this four-generation Pati-Salam model will in fact have two families whose mass is generated perturbatively. It would be interesting to see how non-perturbative effects can generate the Yukawa couplings for the remaining two generations. 

\subsubsection*{Mass terms and net chirality}

Besides Yukawa couplings, discrete flavor symmetries may forbid other kinds of couplings like mass terms between vector-like pairs of of zero modes. In the model at hand such kind of pairs arise in the sector $a_2a_2'$, whose total intersection number is given by $I_{a_2a_2'}=4$. This signals that we have a net chirality of four left-handed chiral multiplets in the representation $(\mathbf 1,\mathbf 1_{+2},\mathbf 1)$, where $\mathbf 1_{+2}$ stands for an antisymmetric representation of $U(2)$.\footnote{More precisely, one computes the spectrum of this sector by applying eqs.(\ref{topintaa'}), with $I_{a_2O6} = 4$. Hence one obtains a net number of four chiral multiplets in the antisymmetric of $U(2)_L$ and no matter in the symmetric representation of $U(2)_L$.} However, this net chirality does not signal the actual content of open string zero modes of this sector. A careful analysis using the rules of subsection \ref{sub:orbichiral} shows that in fact there are nine left-handed chiral multiplets (the ones arising from the wavefunctions of the form (even,odd)) and five right-handed chiral multiplets (the ones from the sector (odd,even)) in the representation $(\mathbf 1,\mathbf 1_{+2},\mathbf 1)$. 

Typically, one would not worry about this mismatch between the zero mode content and the net chiral index, because the ten extra zero modes not accounted by $I_{a_2a_2'}$ naturally arrange into five vector-like pairs that form singlets under the gauge group $U(4) \times U(2)_L \times U(2)_R$. Hence, one expects that the presence of loop corrections or extra compactification ingredients like background fluxes will generate a mass term for these pairs not protected by gauge invariance. 

Nevertheless given a flavor symmetry one needs to check that these pairs of opposite chirality zero modes also form singlets under the discrete flavor group. For the model at hand, table \ref{tab:mass} shows the charges of the different points (or wavefunctions) under the flavor group  (\ref{disZ2Ex4}) for the sector $a_2a_2'$.
\begin{table}[htb] 
\renewcommand{\arraystretch}{1}
 \begin{center}
    \begin{tabular}{ | c || c | c | c |}
    \hline
    Sector & $D_4^{(1)}$ & $D_4^{(2)}$ & $D_4^{(3)}$ \\ \hline\hline
      (even,odd) & $\mathbf 1$ & $\mathbf 1\oplus(+,-)\oplus(-,+)$ & $(+,-)\oplus(-,+)\oplus(-,-)$   \\ \hline
      (odd,even) & $\mathbf 1$ & $(-,-)$ & $\mathbf 1^2\oplus(+,-)\oplus(-,+)\oplus(-,-)$   \\ \hline
      \end{tabular}
\end{center}
\caption{Representations under the dihedral groups of the zero modes in $a_2a_2'$.}
\label{tab:mass}
\end{table}
From there one can see that one cannot form a vector-like pair that is a singlet under the factor $D_4^{(2)}$. As a result, a mass term for any vector-like pair is forbidden by the discrete flavor symmetry. Even if in this particular case the flavor symmetry is approximate, the effect generating such mass term must also break the flavor symmetry (like e.g. non-perturbative effects), and so we expect that such masses for vector-like pairs are  smaller that the ones allowed by all sort of symmetries.

\subsubsection*{The center of the flavor group}

While the flavor symmetry group (\ref{disZ2Ex4}) is non-Abelian, its center can be compared with other discrete Abelian groups present in this model. In particular it can be compared with the $\IZ_N$ discrete gauge symmetries contained in the U(1) factors of  $U(4) \times U(2)_L \times U(2)_R$. These discrete gauge symmetries are discussed in appendix \ref{ap:abe} following the general prescription of \cite{bisu11}. The result is that they are trivial in the sense that they reduce to the center of the gauge group $SU(4)\times SU(2)_L\times SU(2)_R$, which is generated by the elements
\be
g_4=\mbox{diag}(i,i,i,i),\qquad g_{2,L}=\mbox{diag}(-1,-1)_L,\qquad g_{2,R}=\mbox{diag}(-1,-1)_R.
\ee

Let us denote the center of the gauge group by $Z(G)$ and the center of $D_4^{(1)}\times D_4^{(2)}\times D_4^{(3)}$ by $Z(P)$. We would like to know if any subgroup of $Z(P)$ is contained in $Z(G)$ when acting on the Pati-Salam model. Both groups are finite so they have a finite collection of subgroups and this can be answered by direct computation. Table \ref{tab:sub} shows the charges of the visible sector under every $\mathbb Z_2$ subgroup of $Z(P)$ and $Z(G)$.

\begin{table}[htb] 
\renewcommand{\arraystretch}{1}
 \begin{center}
    \begin{tabular}{ | c || c | c | c || c | c | c |}
    \hline
    Sector & $\mathbb Z_2^{(1)}$ & $\mathbb Z_2^{(2)}$ & $\mathbb Z_2^{(3)}$ & $\mathbb Z_{2,C}$ & $\mathbb Z_{2,L}$ & $\mathbb Z_{2,R}$ \\ \hline\hline
      $a_1a_2$ & + & $-$ & + & $-$ & $-$ & +   \\ \hline
      $a_1a_3$ & $-$ & $-$ & + & $-$ & + & $-$   \\ \hline
      $a_2a_3$ & $-$ & + & + & + & $-$ & $-$   \\ \hline
      $a_1a_2'$ & + & + & $-$ & $-$ & $-$ & +   \\ \hline 
      $a_2a_3'$ & $-$ & + & + & + & $-$ & $-$   \\ \hline\hline
      $a_2a_2'$ & + & + & + & + & + & +   \\ \hline 
      $a_3a_3'$ & + & + & + & + & + & +   \\ \hline 
       \end{tabular}
\end{center}
\caption{Charges of the visible sector under $\mathbb Z_2$ subgroups of $Z(P)$ and $Z(G)$.} 
\label{tab:sub}
\end{table}


Looking at Table \ref{tab:sub} we see that $\mathbb Z_2^{(1)}$ and $\mathbb Z_{2,R}$ are the same. Also, the $\mathbb Z_2$ generated by the product of the generators of $\mathbb Z_2^{(3)}$ and $\mathbb Z_{2,C}$ is equivalent to $\mathbb Z_2^{(2)}$ which shows that $\mathbb Z_2^{(2)}$ and $\mathbb Z_2^{(3)}$ are not independent but are related by a gauge transformation. We thus find the discrete flavor group is actually $D_4^{(1)}\times D_4^{(2)}\times D_4^{(3)}/(\mathbb Z_2^{(1)}\times \mathbb Z_2^{(3)})$, and that all the couplings forbidden by this symmetry should be understood in terms of this quotient.

\subsection{A local Pati-Salam three-generation model}

Besides the four-generation model of \cite{bcms05}, one may construct other models in $\IZ_2 \times \IZ_2'$ with semi-realistic spectrum that also display a non-trivial discrete flavor symmetry. In the following we analyze a simple three-generation Pati-Salam model where families transform with non-Abelian representations under a Dihedral flavor group. 

The D6-brane content of the model is shown in table \ref{tab:abc}, where the wrapping numbers $n$, $l$ are arbitrary positive integers. 
\begin{table}[htb]
\renewcommand{\arraystretch}{1}
\begin{center}
	\begin{tabular}{| c | c | c | c |}
	\hline
	$N_\a$ & $(n_\a^1,m_\a^1)$ & $(n_\a^2,m_\a^2)$ & $(n_\a^3,m_\a^3)$ \\ \hline \hline
	$N_a=4$ & $(1,0)$ & $(1,1)$ & $(1,\,$-$1)$ \\\hline
	$N_b=2$ & $(n,\,$-$3)$ & $(0,1)$ & $(3,\,$-$1)$ \\\hline
	$N_c=2$ & $(l,\,$-$1)$ & $($-$2,1)$ & $($-$1,\,$-$1)$ \\ \hline
	\end{tabular}
	\end{center}
\caption{Wrapping numbers for the three-generation Pati-Salam model.}
\label{tab:abc}
\end{table}
Again, this D6-brane content is not sufficient to satisfy the RR-tadpole conditions  (\ref{RRtadpoles}), and extra D-branes should be added in order to construct a complete model. We will then consider it as a local $\IZ_2 \times \IZ_2'$ model, whose discrete flavor symmetry may or may not be broken by the extra D-branes that complete it.

It is easy to see that the gauge group that arises from this D6-brane content is again given by $U(4) \times U(2)_L \times U(2)_R$, and that now the supersymmetry conditions amount to

\be
\begin{array}{c}
U^2\,=\,U^3\\
{\rm tan}^{-1} \left (\frac{3U^1}{n}\right )+{\rm tan}^{-1} \left (\frac{U^3}{3}\right )\,=\,\frac{\pi}{2}\\
{\rm tan}^{-1}\left ( \frac{U^1}{l}\right )+{\rm tan}^{-1}\left ( \frac{U^2}{2}\right )-{\rm tan}^{-1} U^3 \,=\,0
\end{array}
\ee
where $U^i = R_{y_i}/R_{x_i}$. One can solve these equations by setting $n>l>0$, $U^1=\sqrt{\frac{n(n-l)}{2}}$ and $U^2=U^3=\sqrt{\frac{2n}{n-l}}$, hence finding again a $\CN=1$ Pati-Salam model. 

The chiral spectrum of this model can be found by computing the intersection numbers on each two-torus and applying the results of subsection \ref{sub:orbichiral}. The result is displayed in table \ref{tab:sector}, from where it is manifest that all the intersection points in the third torus are even. We then conclude there is a flavor symmetry group of the form
\be
{\bf P}^{abc}_{\T^6/\IZ_2 \times \IZ_2'} \, =\,  D_4
\label{disZ2Ex3}
\ee
where $D_4$ is generated by translations and B-field transformations on $(\T^2)_3$. The zero mode spectrum in the $bc$ sector depends on the integer wrapping numbers $n> l >0$. In the table we have considered the choice $n=2$, $l=1$, which gives a minimal Higgs sector.
\begin{table}[htb] 
\renewcommand{\arraystretch}{1}
 \begin{center}
    \begin{tabular}{ | c || c || c | c | c || c | c |}
    \hline
    Sector & $U(4)\times U(2)_L\times U(2)_R$ &$I_{\a\b}^1$ & $I_{\a\b}^2$ & $I_{\a\b}^3$ &Projection & $I_{\a\b}$ \\ \hline\hline
      $ab$ &$(\bar{\mathbf 4},{\mathbf 2},\mathbf 1)$& -3 & 1 & 2 &$I^1_o\,I^2_e\,I_e^3$ & -2 \\ \hline
      $ab'$ &$(\bar{\mathbf 4},\bar{\mathbf 2},\mathbf 1)$& 3 & -1 & 4 &$I^1_o\,I^2_e\,I_o^3$ & -1  \\ \hline \hline
      $ac$ &$({\mathbf 4},\mathbf 1,\bar{\mathbf 2})$& -1 & 3 & -2 &$I^1_e\,I^2_o\,I_e^3$ & 2  \\ \hline
      $ac'$ &$({\mathbf 4},\mathbf 1,{\mathbf 2})$& -1 & 1 & 0 & $I^1_e\,I^2_e$ & 1  \\ \hline \hline
      $bc$ &$(\mathbf 1,\bar{\mathbf 2},{\mathbf 2})$& $1$ & 2 & -4 &$I^1_e\,I^2_e\,I_o^3$ & -2  \\ \hline 
      $bc'$ &$(\mathbf 1,{\mathbf 2},{\mathbf 2})$ & $5$ & 2 & 2 &$I^1_e\,I^2_e\,I^3_e$ & 12  \\ \hline \hline
      $bb'$ &$(\mathbf 1,\mathbf 1_{+2},\mathbf 1)$& 12 & 0 & 6 &$I^1_e\,I_e^3$ & 18  \\ \hline
       \end{tabular}
\end{center}
\caption{Bulk intersection numbers of the model of table \ref{tab:abc} with $n=2$ and $l=1$, together with the points surviving the orbifold action and the total intersection number.}
\label{tab:sector}
\end{table}

Similarly to the previous example we can easily extract the representation of these chiral Pati-Salam families under the flavor symmetry group $D_4$. We present the result of this analysis in  table \ref{tab:SM}, which shows that in this model one generation is different in the sense that it transforms under an Abelian representation of $D_4$, while the other two form a doublet of the fundamental representation ${\mathbf R_2}$ of the Dihedral group. 

\begin{table}[htb]
\renewcommand{\arraystretch}{1}
\begin{center}
	\begin{tabular}{| c | c | c |}
	\hline
	Sector & Fields & $D_4$ \\ \hline \hline
	$ab$ & $F_R=(\bar{\mathbf 4},{\mathbf 2},\mathbf 1)$ & $\mathbf R_2$ \\\hline
	$ab'$ & $F_R'=(\bar{\mathbf 4},\bar{\mathbf 2},\mathbf 1)$ & $(-,-)$ \\\hline
	$ac$ & $F_L=({\mathbf 4},\mathbf 1,\bar{\mathbf 2})$ & $\mathbf R_2$ \\ \hline
	$ac'$ & $F_L'=({\mathbf 4},\mathbf 1,{\mathbf 2})$ & $(+,+)$ \\ \hline
	$bc$ & $H=(\mathbf 1,\bar{\mathbf 2},{\mathbf 2})$ & $(-,-)\oplus (-,-)$ \\\hline
	$bc'$ & $H'= (\mathbf 1,{\mathbf 2},{\mathbf 2})$& $\stackrel{6}{\oplus}\mathbf R_2$ \\\hline
	\end{tabular}
	\end{center}
\caption{$D_4$ representations.}
\label{tab:SM}
\end{table}
The only Yukawas  allowed by gauge invariance (including anomalous $U(1)$'s) are
\bea
Y:  ab\otimes ac \otimes bc&\longrightarrow&(\bar{\mathbf 4},{\mathbf 2},\mathbf 1)\otimes({\mathbf 4},\mathbf 1,\bar{\mathbf 2})\otimes(\mathbf 1,\bar{\mathbf 2},{\mathbf 2})\\
Y':  ab'\otimes ac\otimes bc'&\longrightarrow&(\bar{\mathbf 4},\bar{\mathbf 2},\mathbf 1)\otimes({\mathbf 4},\mathbf 1,\bar{\mathbf 2})\otimes(\mathbf 1,{\mathbf 2},{\mathbf 2}).
\eea
and under the discrete $D_4$ these coupling behave as
\bea
Y&: & \mathbf R_2\otimes\mathbf R_2\otimes [(-,-)\oplus (-,-)]=\mathbf 1\oplus \mathbf 1\oplus \dots\\
Y'&: & (-,-)\otimes\mathbf R_2\otimes\left (\oplus^6\mathbf R_2\right )=\, \stackrel{6}{\oplus} \mathbf 1\oplus \dots
\eea
where the dots stand for nontrivial representations of $D_4$, and we used $\mathbf R_2\otimes\mathbf R_2=\mathbf 1\oplus(+,-)\oplus(-,+)\oplus(-,-)$. We then conclude that there are a total of eight independent parameters in the Yukawa couplings given by $Y$ and $Y'$.

\section{Conclusions and outlook}
\label{s:conclusions}

In this paper we have analyzed the presence of discrete flavor symmetries in models of intersecting and magnetized D-branes. The general principle to determine the flavor symmetry is to first consider the group ${\bf P}^{\rm bulk}$ of non-trivial metric and B-field transformations that leave the closed string background invariant. In the absence of D-branes, this group of transformations is part of the gauge symmetry group of the 4d effective theory. In the presence of D-branes this gauge group will be partially broken, because the D-brane background is not invariant under its action. The subgroup of isometries and B-field transformations that leave both the closed and open string backgrounds invariant will generate a group of discrete flavor symmetries which in general will be non-Abelian. 

We have implemented the above principle in compactification manifolds like  $\T^6$ and $(\T^2)^3/\IZ_2 \times \IZ_2'$ and orientifolds thereof. In the case of $\T^6$ the initial group of bulk gauge symmetries arising from the metric and B-field is continuous, namely ${\bf P}^{\rm bulk}=U(1)^{12}$, and so one can apply the techniques of \cite{bcmru12} to obtain via dimensional reduction the symmetry group that remains after D-branes have been introduced. The result is a discrete flavor group that fully agrees with the definition in the previous paragraph, as we have checked for models of intersecting and magnetized D-branes. In the case of the $\IZ_2 \times \IZ_2'$ orbifold the bulk gauge group is discrete, namely ${\bf P}^{\rm bulk}=\IZ_2^{12}$ and so we cannot apply the approach of \cite{bcmru12}.\footnote{Using the approach of \cite{bmru13} one shoud be able to embed this discrete bulk gauge group into a continuous one. It would be interesting to also include the presence of D-branes into the formalism of \cite{bmru13} in order to have an alternative derivation of the flavor symmetry group in manifolds with discrete isometries.} However, the strategy followed in this paper does still apply, and so we are able to compute the flavor symmetry group also for this case. The same works for orientifold quotients of the above backgrounds, which in turn allows to study the flavor symmetries of consistent, 4d $\cn=1$ chiral D-brane models like the ones constructed in \cite{bcms05}. We have then analyzed the flavor symmetry group of a couple of semi-realistic Pati-Salam examples, obtaining that the matrices of Yukawa couplings are constrained by the flavor symmetry group beyond the already well-known effect of massive U(1) D-brane symmetries and the Abelian discrete gauge symmetries contained in them \cite{bisu11}.

One of the most attractive features of this approach is its generality, which allows to extend our results in a number of ways. While we have focused on the factorizable orbifold $(\T_2)^3/\IZ_2 \times \IZ_2'$ one can easily generalize our observations to D6-brane models on non-factorizable $\T^6/\IZ_2 \times \IZ_2'$ geometries \cite{fz08}, or to other orbifold geometries like $\T^6/\Z_N$ or $\T^6/\IZ_N \times \IZ_M$ where more realistic D-brane models have been constructed (see e.g. \cite{Honecker:2004kb,Gmeiner:2007zz,Forste:2010gw,Honecker:2012qr}). It would be interesting to see which flavor symmetry groups arise in these other orbifold backgrounds, and in D6-brane models based on smooth Calabi-Yau geometries with discrete isometries\cite{Blumenhagen:2002wn,Uranga:2002pg,Palti:2009bt}. Also while we have considered D-branes that either intersect or carry worldvolume fluxes, one may extend this approach to D-branes that both intersect and  are magnetized, like type IIA models with coisotropic D8-branes \cite{Font:2006na} or type IIB models based on D7-branes (see e.g. \cite{ms04,Blumenhagen:2008zz}). Finally, having a CFT description of the closed string background is not essential in this approach, so one may also extend it  to, e.g., type I compactifications with both open and closed string background fluxes \cite{cm09}.

Another interesting consequence of this approach is that it provides a useful notion of approximate flavor symmetries. Here the principle is again quite simple. If a subgroup ${\bf P}^{abc}$ of the bulk gauge symmetry group ${\bf P}^{\rm bulk}$ leaves a subset A = \{a,b,c\} of three background D-branes invariant, then the sector of the theory given by $U(N_a) \times U(N_b) \times U(N_c)$ will respect the symmetry ${\bf P}^{abc}$, and in particular the tree-level couplings between open string modes of this sector will be invariant under it. If now there is a fourth D-brane $d$ which is not invariant under ${\bf P}^{abc}$ this flavor symmetry group will be broken, and can only be thought as an approximate symmetry of the gauge sector $U(N_a) \times U(N_b) \times U(N_c)$. Nevertheless, in supersymmetric models the holomorphic Yukawa couplings that arise from the subset $A$ of D-branes will be constrained by the discrete symmetry ${\bf P}^{abc}$ at all orders in perturbation theory. Hence, the holomorphic Yukawas forbidden by ${\bf P}^{abc}$ will only be generated at the non-perturbative level, being thus naturally suppressed with respect to the allowed ones. It would be interesting to explore such scenario in specific D-brane models like the ones considered here, in a similar spirit to \cite{a g06,Blumenhagen:2007zk,bcmw12}.

\bigskip

\centerline{\bf \large Acknowledgments}

\bigskip

We would like to thank P.~G.~C\'amara, C.~Hagedorn, L.~E.~Ib\'a\~nez and A.~M.~Uranga for useful discussions. 
This work has been partially supported by the grants FPA2009-07908 and FPA2012-32828 from MINECO, HEPHACOS-S2009/ESP1473 from C.A. de Madrid, the REA grant agreement PCIG10-GA-2011-304023 from the People Programme of FP7 (Marie Curie Action), the grant SEV-2012-0249 of the ÒCentro de Excelencia Severo OchoaÓ Programme and the SPLE Advanced Grant under contract ERC-2012-ADG 20120216-320421. F.M. is supported by the Ram\'on y Cajal programme through the grant RYC-2009-05096. D.R. is supported through the FPU grant AP2010-5687. L.V. is grateful for support from CONACyT and to the IFT-UAM/CSIC for hospitality.

\clearpage

\appendix


\section{The $\IZ_2 \times \IZ_2'$ orbifold}
\label{ap:review}

Let us consider type IIA string theory compactified on the toroidal orbifold background $\T^6/(\IZ_2\times \IZ_2)$, the $\IZ_2$  generators acting as
\be
\Theta:
\left\{
\begin{array}{l}
\vspace*{-.25cm}
z_1\to -z_1 \\ \vspace*{-.25cm}
z_2\to -z_2 \\ z_3\to z_3 
\end{array}\right.
\quad\quad\quad\quad
\Theta':
\left\{
\begin{array}{l}
\vspace*{-.25cm}
z_1\to z_1 \\ \vspace*{-.25cm}
z_2\to -z_2 \\ z_3\to -z_3 
\end{array}\right.
\label{ap:orbiaction}
\ee 
on the three complex coordinates of $\T^6 = (\T^2)_1 \times (\T^2)_2 \times (\T^2)_3$. Besides such action one must specify the choice of discrete torsion that relates these two $\IZ_2$ group generators. As explained in \cite{vafa86} there are two inequivalent choices, whose twisted homologies are either $(h_{11}^{\rm tw.}, h_{21}^{\rm tw.}) = (48,0)$ or $(h_{11}^{\rm tw.}, h_{21}^{\rm tw.}) = (0,48)$. Similarly to \cite{bcms05} we will consider the second case, and dub it as $\IZ_2\times \IZ_2$ orbifold with discrete torsion or $\IZ_2\times \IZ_2'$. Such background then contains 96 collapsed three-cycles at the fixed loci of (\ref{ap:orbiaction}). 

\begin{figure}[htb]
  \centering
  \def\svgwidth{490pt}
  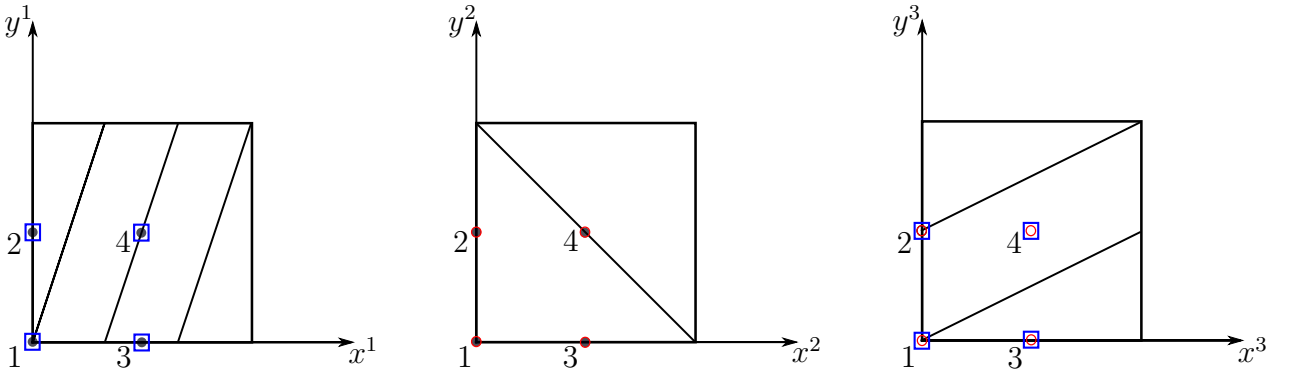
  \caption{Fractional brane passing through 4 fixed points for each twisted sector. Fixed points are denoted by dots in the $\Theta$ sector, by circles in the $\Theta'$ sector and squares in the $\Theta\Theta'$ sector.}
  \label{fig:fixed}
\end{figure}

Let us now add space-time filling D6-branes wrapping supersymmetric three-cycles of this toroidal orbifold. In terms of a factorized $\T^6$ geometry these can be described as the product of three one-cycles
\be
\label{ap:torcycle}
[\Pi_a]=\bigotimes_{i=1}^3 \left( n^i_a\ [a^i] +  m^i_a\ [b^i] \right) \quad \quad n^i_a, m^i_a \in \IZ {\rm \ and \ coprime}
\ee
where $[a^i]$, $[b^i]$ are the homology classes of the fundamental one-cycles of $(\T^2)_i$. Notice that the $\T^6$ homology class $[\Pi_a]$ is invariant under the orbifold action (\ref{ap:orbiaction}), and so one can consider three-cycle representatives $\Pi_a$ also invariant under (\ref{ap:orbiaction}). A D6-brane wrapping an invariant three-cycle will suffer the orbifold projection on its Chan-Paton degrees of freedom, resulting into fractional D6-branes with non-vanishing charge under the RR twisted sector of the theory. Geometrically, on each $(\T^2)_i$ a fractional D6-branes goes through two fixed points of the action $z_i \to -z_i$, and it wraps collapsed three-cycles that correspond to such fixed points (see fig. \ref{fig:fixed}). Precisely because of this, fractional D6-branes are `rigid': they cannot be taken away from a fixed locus of the action (\ref{ap:orbiaction}), and so they do not contain the deformation moduli typical of D-branes in toroidal compactifications.

Following \cite{bcms05} the homology class of a fractional D6-brane is of the form
\be 
\label{rigid}
[\Pi^F_a]\, =\, {1\over 4}\, [\Pi^B_a] +
{1\over 4} \left( \sum_{I,J \in S_\Theta^a} \epsilon^\Theta_{a,IJ}\,  
[\Pi^\Theta_{IJ,\,a}] \right)+
 {1\over 4} \left( \sum_{J,K\in S_{\Theta'}^a} \epsilon^{\Theta'}_{a,JK}\, 
[\Pi^{\Theta'}_{JK,\,a}] \right)+
 {1\over 4}\left( \sum_{I,K\in S_{\Theta\Theta'}^a} 
\epsilon^{\Theta\Theta'}_{a,IK}\, [\Pi^{\Theta\Theta'}_{IK,\,a}] \right)
\ee
where $[\Pi_a^B]$ stands for a {\em bulk} three-cycle, that is a $\T^6$ three-cycle of the form (\ref{ap:torcycle}) that is inherited in the orbifold quotient. Bulk three-cycles correspond to the untwisted RR charges of the orbifold, and the intersection number between them is given by
\be
\label{prod}
I_{ab}^B =
[\Pi_a^B] \cdot [\Pi_b^B] = 
4\prod_{i=1}^3 (n^i_a\, m^i_b - m^i_a\, n^i_b),
\ee
where the factor of 4 arises from taking into account the $\IZ_2 \times \IZ_2'$ orbifold action. Beside the bulk cycles there are 32 collapsed three-cycles for each of the three twisted sectors. Their homology class is of the form
\be
[\Pi^g_{IJ,\,a}]= 2 [e_{IJ}^g] \otimes \left(n_a^{i_g} [a^{i_g}] +  m_a^{i_g} [b^{i_g}]\right) 
\ee
where $g = \Theta, \Theta', \Theta\Theta'$ runs over all  twisted sectors, and $i_g = 3,1,2$, respectively. Here $e_{IJ}^g$ $I,J\in\{1,2,3,4\}$ stand for the 16 fixed points on $(\T^2)_i \times (\T^2)_j/\IZ_2$, where $\IZ_2 = \{ 1, g\}$ and $(\T^2)_{i,j}$ are the two-tori such that $g: (z_i, z_j) \mapsto (-z_i, -z_j)$ (see fig. \ref{fig:fixed}). These fixed points correspond to the $\IZ_2$ singularities of a {\bf K3} surface in its orbifold limit $\T^4/\IZ_2^2$, and each can be blown up to a ${\bf P}^1$ whose homology class is given by $[e_{IJ}^g]$. Finally, $a^{i_g}$, $b^{i_g}$ stand for the fundamental one-cycles of $(\T^2)_{i_g}$, the two-torus which is left invariant under the action of $g$. Gathering all these facts together, one can compute the intersection number of two collapsed three-cycles as 
\be 
\label{inttwist} 
[\Pi^g_{IJ,\,a}] \cdot [\Pi^h_{KL,\,b}]\,
=\, 4\, \d_{IK} \d_{JL} \d^{gh} \,
(n_a^{i_g}\, m_b^{i_g} - m_a^{i_g}\, n_b^{i_g})
\ee

The homology class (\ref{rigid}) is given by a particular linear combination of bulk and collapsed three-cycles, which is determined as follows. In the covering space $(\T^2)^3$ a BPS D6-brane looks like as a product of three 1-cycles with wrapping numbers $(n_a^i, m_a^i)$ and constant slope, see figure \ref{fig:fixed}. Fractional D6-branes must be invariant under (\ref{ap:orbiaction}), and so on each two-torus they must pass through two fixed points of $(\T^2)_i/\IZ_2$ with $\IZ_2 = \{1, z_i \mapsto -z_i\}$. Which are these fixed points depends on the wrapping numbers $(n_a^i, m_a^i)$, see table \ref{fixed} in the main text.

Let us now consider a particular twisted sector, say $g =\Theta$. The collapsed three-cycles of this sector are related to the fixed points $e_{IJ}^\Theta$ of $(\T^2)_1 \times (\T^2)_2/\{1, \Theta\}$. A fractional D6-brane will pass through 4 fixed points $e_{IJ}^\Theta$. More precisely, the index $I$ will take two different values specified by $(n_a^1, m_a^1)$ and one of the choices in table \ref{fixed}, while $J$ will be constrained by $(n_a^2, m_a^2)$. This subset of $2 \times 2$ elements $\{(I,J)\} \subset \{1,2,3,4\} \times \{1,2,3,4\}$ is denoted as $S_\Theta^a$ in (\ref{rigid}), and similar definitions apply to $S_{\Theta'}^a$ and $S_{\Theta\Theta'}^a$. It is easy to see that given the bulk wrapping numbers $(n_a^i, m_a^i)$, $i=1,2,3$ there are eight different choices for specifying $S_\Theta^a$, $S_{\Theta'}^a$ and $S_{\Theta\Theta'}^a$. From the viewpoint of the covering space $(\T^2)^3$, these choices correspond to the  $2^3$ different locations that an invariant three-cycle can have.

Besides $S_g^a$ one needs to specify the signs $\epsilon^\Theta_{a,IJ},\,\epsilon^{\Theta'}_{a,JK},\,\epsilon^{\Theta\Theta'}_{a,IK}\,=\,\pm 1$ that appear in (\ref{rigid}). These signs are not arbitrary but must fulfill several consistency conditions discussed in \cite{bcms05}. One finds that there are essentially 8 inequivalent choices for these signs. In general the set of fixed points is given by
\be
\begin{array}{rcl}
S_{\Theta} & = & \left\{ \{I_1, I_2\} \times \{J_1, J_2\} \right\} \\
S_{\Theta'} & = & \left\{ \{J_1 J_2\} \times \{K_1 K_2\} \right\} \\
S_{\Theta\Theta'} & = & \left\{ \{K_1 K_2\} \times \{I_1 I_2\} \right\}
\end{array}
\label{twistedsets}
\ee
where $i_\a, j_\a, k_\a$, $\a =1,2$ represent fixed point coordinates in the first, second and third $\T^2$ factors, respectively. If we fix $\eps_{I_1J_1}^{\Theta} = \eps_{J_1K_1}^{\Theta'} = \eps_{K_1I_1}^{\Theta\Theta'} = +1$ then all the other $\eps$'s depend on only three independent signs. More precisely we have that $\eps^\Theta_{I_2J_1} = \eps^{\Theta\Theta'}_{K_1I_2} = \eps_I$, $\eps^\Theta_{I_1J_2} = \eps^{\Theta'}_{J_2 K_1} = \eps_J$, $\eps^{\Theta'}_{J_1K_2} = \eps^{\Theta\Theta'}_{K_2 I_1} = \eps_K$ and $\eps^\Theta_{I_2 J_2} = \eps_I \eps_J$, $\eps^{\Theta'}_{J_2 K_2} = \eps_J \eps_K$, $\eps^{\Theta\Theta'}_{K_2 I_2} = \eps_K \eps_I$, with $\eps_I, \eps_J, \eps_K = \pm 1$. The choice of these three signs can be interpreted as the choice of discrete Wilson lines for a fractional D6-brane along each one-cycle. 

Having fixed $S_g^a$ and $\eps_{a,IJ}^{\, g}$ as above, there are four inequivalent choices of wrapping numbers $(n_a^I, m_a^I)$ which correspond to the same bulk three-cycle $\Pi_a^B$ but to different fractional three-cycle $\Pi_a^F$. These are given by
\be
\begin{array}{ccc}
(n_a^1,m_a^1)   & (n_a^2,m_a^2)  & (n_a^3,m_a^3) \\
(-n_a^1,-m_a^1)  & (-n_a^2,-m_a^2)  & (n_a^3,m_a^3) \\
(n_a^1,m_a^1)  & (-n_a^2,-m_a^2)  & (-n_a^3,-m_a^3) \\
(-n_a^1,-m_a^1)  & (n_a^2,m_a^2)  & (-n_a^3,-m_a^3) \\
\end{array}
\label{ineqbulk}
\ee
and can be interpreted as the four different $\IZ_2 \times \IZ_2'$ twisted charges that a fractional D6-brane can have. Indeed, it is easy to see that one obtains a pure bulk D6-brane by adding these four fractional D6-branes. One can further support this claim by computing the chiral spectrum between two fractional D6-branes, as we now proceed to show.

\subsubsection*{Chiral index}

Given two stacks of fractional D6-branes wrapped on $\Pi_a^F$ and $\Pi_b^F$ one can easily compute the chiral spectrum of open strings with one endpoint on each of them. Indeed, let us consider $N_a$ D6-branes wrapped on $\Pi_a^F$ and $N_b$ D6-branes on $\Pi_b^F$. Then the chiral spectrum will be given by $I_{ab}$ left-handed chiral multiplets in the bifundamental $(N_a, \bar{N}_b)$ representation of $SU(N_a) \times SU(N_b)$. Here $I_{ab} = [\Pi_a^F] \cdot [\Pi_b^F]$ is the topological intersection number of the two three-cycles, and can be computed from (\ref{prod}), (\ref{inttwist}) and the fact that an intersection number between a bulk  and a collapsed three-cycle vanishes.

For instance, let us consider the case where $\Pi_a^F$, $\Pi_b^F$ are such that they have trivial discrete Wilson lines ($\eps_{a,b}^{\, g} = 1$ in (\ref{rigid})) and they both intersect the origin of $(\T^2)^3$ ($I_1=J_1=K_1=1$ in (\ref{twistedsets}) for $S_g^{a,b}$). Then the intersection number $I_{ab}$ is specified by the bulk wrapping numbers
\be
\begin{array}{lcccc}
\Pi^F_{a} & : & (n_a^1, m_a^1) &  (n_a^2, m_a^2)  & (n_a^3, m_a^3) \\
\Pi^F_{b} & : & (n_b^1, m_b^1) & (n_b^2, m_b^2) & (n_b^3, m_b^3) 
\end{array}
\label{ap:D6ab}
\ee
%
More precisely we find that
\be
I_{ab} = [\Pi^F_{a}] \cdot [\Pi^F_{b}]  =  \frac{1}{4} \left[ I_{ab}^1I_{ab}^2I_{ab}^3 +   I_{ab}^1\, \rho_2\rho_3  +  I_{ab}^2\, \rho_1\rho_3 +  I_{ab}^3 \, \rho_1\rho_2 \right] 
\label{ap:topint}
\ee
where $I_{ab}^i = n_a^im_b^i - n_b^im_a^i$ and  $\rho_i$ is defined as in (\ref{rho}).
Despite the factor of $1/4$ one can check that such intersection number is always an integer, as required by consistency. Notice that the bulk intersection number $I_{ab}^B$ remains unchanged if we replace $\Pi_a^F$ by any of the other bulk wrapping numbers in (\ref{ineqbulk}). The intersection numbers $I_{ab}^i$ for each individual two-torus do however depend on this choice, and so does the total intersection number $I_{ab}$. As show in the main text this formula is reproduced by considering those linear combinations of intersection points invariant under the orbifold action. As one can also see from that discussion, the four different type of projection that depend on the signs $s_1$, $s_2$, $s_3$ can be obtained by considering the pair of D6-branes (\ref{ap:D6ab}) and then replacing $\Pi_a^F$ by any of the other bulk wrapping numbers in (\ref{ineqbulk}). This shows in more detail that each of these D6-branes has a different Chan-Paton factor, because for the same bulk embedding the open strings ending in $\Pi_a$ feel a different orbifold action, adding up to the regular representation of $\IZ_2 \times \IZ_2'$.


\section{Flavor symmetries from dimensional reduction}
\label{ap:inter}

Let us consider the dimensional reduction that yields the St\"uckelberg lagrangians (\ref{st1}-\ref{st2}) in 4d for D6-branes at angles in type IIA on $\T^2\times \T^2\times \T^2$ which shows the appearance of discrete symmetries. In \cite{bcmru12} this was done for a toroidal compactification of Type I with magnetic fluxes. In our case we should consider the Type IIA supergravity together with the DBI action for the branes at angles to get the full non-Abelian structure. We will, however, take a simpler approach and consider only the DBI part to derive the abelian part of the symmetry.

Consider a D6-brane wrapping a factorizable three-cycle $\Pi_a=(n_a^1,m_a^1)\otimes(n_a^2,m_a^2)\otimes(n_a^3,m_a^3)$. The DBI action for such a D6-brane is
\be\label{dbi}
S_6=-\mu_6\int_{M_4}d^{4}x\int_{\Pi_a}d^3q\, e^{-\Phi}\sqrt{-\det(P[G]+P[B]-kF)}
\ee
with $k=2\pi\alpha'$ and $P[\cdot]$ is the pullback on the worldvolume of the brane which looks like
\be
P[A]_{\a\b}=A_{\a\b}+A_{ij}\p_\a\phi^i\p_\b\phi^j+\p_\a\phi^iA_{i\b}+\p_\b\phi^iA_{\a i}
\ee
where $\a,\b$ are indices on the brane and $i,j$ are transverse. $\phi^i$ are the embedding functions of the brane in the bulk. Using the Taylor expansion of the determinant
\be
\det (1+M)=1+\tr\, M+\frac{1}{2}[\tr \,M]^2-\frac{1}{2}\tr \,M^2+\dots
\ee
we can expand the action (\ref{dbi}) in derivatives. Namely,
\bea\nonumber
S_6&=&-\mu_6\int_{M_4}d^{4}x\int_{\Pi_a}d^3q\, e^{-\Phi_0}\sqrt{-\det G_{\a\b}}\left ( 1+\frac{1}{2}G^{\a\b}G_{ij}\p_\a\phi^i\p_\b\phi^j+G^{\a\b}\p_\a\phi^iG_{i\b}  \right. \\\label{dbi2}
&&\hspace{6.5cm}  \left. -\,\frac{k}{2}B_{\a\b}F^{\a\b}+\frac{k^2}{4}F_{\a\b}F^{\a\b}+\dots \right )
\eea
where we only kept the terms quadratic in fluctuations. Since the brane is wrapping the cycle $\Pi_a$ we take the following rotated coordinates in $\T^2\times \T^2\times \T^2$
\be
q^l=x^l\cos \th_l+y^l\sin \th_l,\qquad p^l=-x^l\sin\th_l+y^l\cos\th_l,\qquad\quad \tan\th_l=\frac{m_a^l}{n_a^l}
\ee
with $x^l,\,y^l$ real coordinates on $(\T^2)_l$ for $l=1,2,3$, the $q^l$'s are along the brane and the $p^l$'s transverse to it. Going back to the action (\ref{dbi2}) we get the following terms 
\bea\nonumber
S_6&\supset&-\mu_6\int_{M_4}d^{4}x\int_{\Pi_a}d^3q\, e^{-\Phi_0}\sqrt{-\det G_{\a\b}}     \left (\frac{1}{2}G^{\mu\nu}G_{pp}\p_\mu\phi^p\,\p_\nu\phi^p+G^{\mu\nu}\p_\mu\phi^p\, G_{p\nu}  \right. \\\label{dbi3}
&&\hspace{6.5cm}  \left. -\, kB_{\mu q}F^{\mu q}+\frac{k^2}{2}F_{\mu q}F^{\mu q}+\dots \right ).
\eea
In this expression the indices $\mu,\nu$ are in 4d, while $p$ and $q$ run through $p^l$ and $q^l$ respectively. The first line yields\footnote{The kinetic term of the gauge bosons $V_{\mu}^x$ and $V_\mu^y$ that complete the St\"uckelberg Lagrangian can be obtained from dimensional reduction of the closed string sector of the theory.}
\be
\CL_{\rm St}\, =\, -\frac{1}{2}\sum_{i=1}^3 \left (\p_\mu\phi_a^i-m_a^i V_{\mu}^{x_i} +n_a^i V_{\mu}^{y_i} \right )^2
\ee
where we defined $\phi_a^i= \sqrt{n_i^2+m_i^2}\,\phi^i$ so that $\phi_a^i\sim\phi_a^i+1$ following the conventions in \cite{bcmru12}. This is the Lagrangian (\ref{st1}) that describes the spontaneous breaking of the continuous isometry group $U(1)^6$ of the torus to $U(1)^3\times\mathbb Z_{q_1}\times\mathbb Z_{q_2}\times\mathbb Z_{q_3}$ with $q_i=(n_a^i)^2+(m_a^i)^2$ due to the presence of the brane.\footnote{See section 2.5 in \cite{bcmu12} for a discussion on the discrete part of this group.} Also, $\phi_a^i$ provide the longitudinal degree of freedom to the massive gauge bosons $-m_a^i V_{\mu}^{x_i} +n_a^i V_{\mu}^{y_i}$.

Furthermore, from the second line in (\ref{dbi3}) one finds the following contribution to the low energy action
\be
\CL_{\rm St}\, =\, -\frac{1}{2}\sum_{i=1}^3 \left (\p_\mu\xi_a^i-n_a^i B_{\mu}^{x_i} - m_a^i B_{\mu}^{y_i} \right )^2
\ee
where $\xi_a^i=\sqrt{n_i^2+m_i^2}\,A^{i}$ and we have rescaled the B-field as $B\rightarrow k^{-1} B$, c.f.(\ref{st2}). 

This analysis shows that the presence of a single brane breaks the continuous $U(1)^{12}$ gauge symmetry that arises from the reduction of the metric and B-field down to $U(1)^6$ plus some discrete part.\footnote{In case that we have an orientifold background the bulk symmetry is not $U(1)^{12}$ but $U(1)^6 \times \IZ_2^6$. Indeed, because the O6-planes are located along $y^i=0,\frac{1}{2}$ for $i=1,2,3$, the U(1) symmetries generated by $V_{\mu}^{y_i}$ and $B_{\mu}^{x_i}$ are broken down to $\mathbb Z_2$, obtaining a residual $\mathbb Z_2\times \mathbb Z_2$ gauge group for each $(\T^2)_i$. \label{footori}} It is clear that adding more branes will generically break the gauge symmetry completely. Indeed, the Lagrangian for a set of intersecting branes will include the terms
\be\label{stt}
\CL_{\rm St}\, =\, -\frac{1}{2}\sum_{\a}\sum_{i=1}^3\left \{ \left (\p_\mu\phi_\a^i-m_\a^i V_{\mu}^{x_i} +n_\a^i V_{\mu}^{y_i} \right )^2+ \left (\p_\mu\xi_\a^i-n_\a^i B_{\mu}^{x_i} - m_\a^i B_{\mu}^{y_i} \right )^2 \right \}
\ee
where $\a$ runs over the branes. Unless all the branes are parallel in a given torus this will Higgs the continuous part of the gauge group completely. Nevertheless, there can be a discrete remnant which we discuss in the following.

Let us restrict to the case where there are only two branes $a$ and $b$ and focus on the part of the action (\ref{stt}) that involves $V_{\mu}^{x_i},\,V_{\mu}^{y_i}$. Following \cite{bcmru12}, one can see that the discrete gauge group coming from the $i$-th torus is
\be
\mathcal T^{ab}_i=\frac{\G_i}{\hat \G_i}
\ee
where $\G_i$ is the lattice of the $i$-th torus and $\hat \G_i$ is the lattice generated by the intersection points. Namely,
\be
\G_i=\langle(1,0),(0,1) \rangle,\qquad\hat \G_i= \frac{1}{I_{ab}^i}\langle(n_a^i,m_a^i),(n_b^i,m_b^i) \rangle.
\ee
One can check that indeed $\mathcal T^{ab}_i=\mathbb Z_{I_{ab}^i}$ and since the three-cycles the D6-branes wrap are factorizable we have
\be
{\bf \ct}^{ab}_{\T^6} \, =\, \IZ_{I_{ab}^1} \times  \IZ_{I_{ab}^2} \times  \IZ_{I_{ab}^3}
\ee
which reproduces eq.(\ref{disT6}) in the main text. A completely analogous argument shows that the second term in (\ref{stt}) yields
\be
{\bf \cw}^{ab}_{\T^6} \, =\, \IZ_{I_{ab}^1} \times  \IZ_{I_{ab}^2} \times  \IZ_{I_{ab}^3}
\ee
in agreement with eq.(\ref{disT6W}).

These two groups, ${\bf \ct}^{ab}_{\T^6}, \,{\bf \cw}^{ab}_{\T^6}$, do not commute as can be seen from their action on the wavefunctions of chiral matter in the $ab$ sector. Instead they generate the non-Abelian discrete group
\be
{\bf P}^{ab}_{\T^6} \, =\, H_{I_{ab}^1} \times  H_{I_{ab}^2} \times  H_{I_{ab}^3}
\ee
with $H_N \simeq ( \IZ_N \times \IZ_N ) \rtimes \IZ_N$.

\subsubsection*{Magnetized D-branes}

In order to connect with the dimensional reduction of Type I with magnetized D9-branes performed in section 6.2 of \cite{bcmru12} we T-dualize the above setup along the directions $y^i$ for $i=1,2,3$. Thus, as usual D6 branes at angles turn into D9 branes with magnetic fluxes given by
\be
F^a_{x^iy^i}=\frac{m_a^i}{kn_a^i}\mathbb I_{n^i_a}.
\ee
Notice that since the three-cycles the D6-branes wrap are factorizable the `nondiagonal' fluxes $F_{z^i\bar z^j}$ are zero for $i\neq j$. This means that just like before the dimensional reduction on $\T^6$ factorizes in $(\T^2)_1\times (\T^2)_2\times (\T^2)_3$. More precisely, the above computation gives in this case
\be\label{d9s}
\CL_{\rm St}\, =\, -\frac{1}{2}\sum_{\a}\sum_{i=1}^3\left \{ \left (\p_\mu\xi_{x,\a}^i- n_\a^iB_{\mu}^{x_i} -m_\a^i V_{\mu}^{y_i} \right )^2+ \left (\p_\mu\xi_{y,\a}^i+m_\a^i V_{\mu}^{x_i} -n_\a^i B_{\mu}^{y_i} \right )^2 \right \}
\ee
where the axions $\xi_{x,\a}^i,\,\xi_{y,\a}^i$ correspond to the Wilson lines on the worldvolume of the brane along $x^i$ and $y^i$ respectively and have periodic identifications $\xi_{q,\a}^i\sim\xi_{q,\a}^i+1$. In order to cancel the total D9-charge we must include an orientifold projection acting trivially on the tori which introduces O9-planes. Also, the B-field does not survive the projection so we can safely set it to zero\footnote{Actually, a subgroup $\IZ_2^6$ survives the orientifold projection, as in the type IIA case (see footnote \ref{footori}).} in (\ref{d9s}) which reproduces the result in \cite{bcmru12}, namely,
\be\label{12}
\CL_{\rm St}\, =\, -\frac{1}{2}\sum_{\a}\sum_{i=1}^3\left \{ \left (\p_\mu\xi_{x,\a}^i -\frac{m_\a^i}{n_\a^i} V_{\mu}^{y_i} \right )^2+ \left (\p_\mu\xi_{y,\a}^i+\frac{m_\a^i}{n_\a^i} V_{\mu}^{x_i}\right )^2 \right \}
\ee
with $\xi_{q,\a}^i\sim\xi_{q,\a}^i+ 2/n_\a^i$.\footnote{The factor 2 appears due to the orientifold projection.}


\section{Abelian discrete gauge symmetries in $\IZ_2 \times \IZ_2'$}
\label{ap:abe}

In this appendix we describe a different kind of discrete gauge symmetries that arise for models of D-branes on the $\IZ_2 \times \IZ_2'$ orientifold. As shown in \cite{bisu11}, massive U(1) D-brane symmetries have $\IZ_k$ subgroups that survive as discrete gauge symmetries in the low energy effective action. Even if they are Abelian and typically flavor blind, one should take them into account in order to describe the full 4d discrete gauge group of a given D-brane model. In the following we will apply the general discussion of \cite{bisu11} to the particular case of type IIA models in the $\IZ_2 \times \IZ_2'$ orientifold. We refer the reader to \cite{Honecker:2013hda} for a more detailed discussion of these symmetries in the context of toroidal orientifolds.

Let us consider type IIA orientifold compactifications with D6-branes. Following \cite{cim11}, we may take a linear combination of the D-brane $U(1)$ symmetries and associate a three-cycle $\Pi_\a$ of the manifold to the resulting $U(1)_\a$, which will be a linear combination of the three-cycles wrapped by the D6-branes of the model. A $U(1)_\a$ is massless if
\be
[\Pi_\a^-] \equiv [\Pi_\a] - [\Pi_\a'] = 0
\ee
That is, if $\Pi_\a$ minus its  orientifold image $\Pi_\a'$ is trivial in homology. All the other U(1)'s are massive, and will be broken by D2-brane instanton effects. Nevertheless, a $\IZ_k$ gauge symmetry will remain if the intersection numbers of all possible D2-instantons with $\Pi_\a^-$ are a multiple of $k$. In practice that amounts to check that \cite{bisu11}
\be
[\Pi_\a] \cdot ([\Pi_\b] + [\Pi_{\b}'])\, =\, 0 \ {\rm mod}\ k
\label{modk}
\ee
for any three-cycle $\Pi_\b$ of the compactification manifold. For the case of the $\IZ_2 \times \IZ_2'$ orientifold, this means that $\Pi_\b$ runs over all possible fractional three-cycles of the form (\ref{rigid}), that we can represent as 
\be 
\label{rigid2}
\Pi^F_\b\, =\,  (n_\b^1,m_\b^1) \  (n_\b^2,m_\b^2)  \  (n_\b^3,m_\b^3)
\ee
and then we must remember for each choice (\ref{rigid2}) there are 8 different choices of discrete positions (sets of fixed points where the three-cycles can go through) and 8 different choices of discrete Wilson lines (choices for $\epsilon^g_{ij,\,a}$). In terms of the action of the orientifold, fractional three-cycles can be classified as follows:

\subsubsection*{Fully invariant three-cycles}

These are the fractional three-cycles of the form:
\be
\begin{array}{c}
(1,0) \ (1,0) \ (1,0) \\
(1,0)  \ (-1,0) \ (-1,0) \\
 (-1,0) \  (1,0) \ (-1,0) \\
 (-1,0)  \ (-1,0) \ (1,0) \\
\end{array}
\label{inv}
\ee
that are fully invariant under the orientifold action. In this case the orientifold projection is such that a D2-brane instanton wrapping these cycles will develop a gauge group projected down to $O(1)$, and so we should not include $\Pi_\b'$ in (\ref{modk}).  For each of these four fractional three-cycles we have 64 possibilities for their fractional positions and Wilson lines. The computation of the intersection number between a fractional brane 
\be
N_a \Pi_a^F \, =\, N_a (n_a^1,m_a^1) \  (n_a^2,m_a^2)  \  (n_a^3,m_a^3)
\label{exagen}
\ee
and the three-cycles (\ref{inv}) is quite similar to the discussion of Appendix A of \cite{bcms05}. The result is then that the intersection numbers are of the form 
\be
\begin{array}{l}\vspace*{.2cm}
 {1 \over 4} N_a \left[ m_a^1 m_a^2 m_a^3 + Q_1^a m_a^1 + Q_2^a m_a^2 + Q_3^a m_a^3 \right] \\ \vspace*{.2cm}
{1 \over 4} N_a \left[ m_a^1 m_a^2 m_a^3 + Q_1^a m_a^1 - Q_2^a m_a^2 - Q_3^a m_a^3 \right] \\ \vspace*{.2cm}
 {1 \over 4} N_a \left[ m_a^1 m_a^2 m_a^3 - Q_1^a m_a^1 + Q_2^a m_a^2 - Q_3^a m_a^3 \right] \\ \vspace*{.2cm}
 {1 \over 4} N_a \left[ m_a^1 m_a^2 m_a^3 - Q_1^a m_a^1 - Q_2^a m_a^2 + Q_3^a m_a^3 \right] 
\end{array}
\label{Ktheory}
\ee
with $Q_a^i$ integer numbers that depend on the specific three-cycle $\Pi_a^F$ that the D-brane is wrapping. More precisely,
\be
Q_i^a\, =\, \left\{
\begin{array}{ccl}\vspace*{.1cm}
1 & {\rm if} & m_a^j m_a^k \equiv 1\, {\rm mod}\, 2 \\\vspace*{.1cm}
2\ {\rm and}\ 0 & {\rm if} & m_a^j m_a^k \equiv 0 \, {\rm mod} \, 2 \quad {\rm and} \quad  m_a^j+m_a^k \equiv 1\,   {\rm mod}\, 2 \\
4\ {\rm and}\ 0 & {\rm if} & m_a^j m_a^k \equiv 0 \,  {\rm mod}\, 2 \quad {\rm and} \quad  m_a^j+m_a^k =0 \,   {\rm mod}\, 2 
\end{array}
\right.
\label{Qs}
\ee

For instance, let us take the fractional three-cycle
\be
\Pi^F_{a_2}\, =\, (1,0)  \  (2,1)  \  (4,-1) 
\label{exa2}
\ee
Then we have that the intersection numbers are given by (\ref{Ktheory}) with
\be
Q_2^{a_2} = Q_3^{a_3} = 0 \quad {\rm or} \quad Q_2^{a_2} = Q_3^{a_3} = 2 
\ee
in agreement with the observations above. The intersection numbers with the fully invariant three-cycles (\ref{inv}) are then
\be
0, \pm 1
\ee
so that we already know that the g.c.d. of the intersection numbers is 1 and no discrete gauge symmetry remains except the center of the gauge group $SU(N_{a_2})$.

\subsubsection*{Bulk invariant three-cycles}

There are some other kind of fractional three-cycles whose bulk piece $\Pi^B$ is invariant under the action of the orientifold but that is not true for its fractional three-cycle content. One example is given by
\be
(1,0) \ (0,1) \ (0,-1) \quad \stackrel{\Omega{\cal R}}{\longrightarrow} \quad (1,0) \ (0,-1)  \ (0,1)
\ee
Hence, we need to consider invariant combinations $[\Pi_\b] + [\Pi_\b']$ of these three-cycles, namely
\be
\begin{array}{c}
\, [(1,0) \  (0,1)  \ (0,-1)] \ + \ [(1,0)  \ (0,-1)  \ (0,1)] \\ 
\, [(-1,0) \  (0,1) \ (0,1)] + [(-1,0)  \ (0,-1)  \ (0,-1)] \\ 
\, [(0,1) \  (1,0)  \ (0,-1)] \ + \ [(0,-1)  \ (1,0)  \ (0,1)] \\  
\, [(0,1) \  (-1,0)  \ (0,1)] \ + \ [(0,-1)  \ (-1,0)  \ (0,-1)] \\ 
\, [(0,-1) \  (0,1)  \ (1,0)] \ + \ [(0,1)  \ (0,-1)  \ (1,0)] \\ 
\, [(0,1) \  (0,1)  \ (-1,0)] \ + \ [(0,-1)  \ (0,-1)  \ (-1,0)] \\
\end{array}
\label{bulkinv}
\ee
The intersection numbers that we obtain with (\ref{exagen}) are
\be
\begin{array}{l}\vspace*{.2cm}
{1 \over 2} N_a \left[ - m_a^1 n_a^2 n_a^3 \pm Q_1^a m_a^1 \right] \\ \vspace*{.2cm}
{1 \over 2} N_a \left[ - n_a^1 m_a^2 n_a^3 \pm Q_2^a m_a^2 \right] \\ \vspace*{.2cm}
{1 \over 2} N_a \left[ - n_a^1 n_a^2 m_a^3 \pm Q_3^a m_a^3 \right]
\end{array}
\ee
where
\be
Q_i^a\, =\, \left\{
\begin{array}{ccl}\vspace*{.1cm}
1 & {\rm if} & n_a^j n_a^k \equiv 1\, {\rm mod}\, 2 \\\vspace*{.1cm}
2\ {\rm and}\ 0 & {\rm if} & n_a^j n_a^k \equiv 0 \, {\rm mod}\, 2 \quad {\rm and} \quad  n_a^j+n_a^k \equiv 1\,  {\rm mod}\, 2 \\
4\ {\rm and}\ 0 & {\rm if} & n_a^j n_a^k \equiv 0 \, {\rm mod}\, 2 \quad {\rm and} \quad  n_a^j+n_a^k =0 \,  {\rm mod}\, 2 
\end{array}
\right.
\label{Qs2}
\ee
considering all the possibilities for the discrete positions and Wilson lines. For instance, taking again the example (\ref{exa2}) we have that
\be
Q_2^{a_2} =  0, 2 \quad {\rm and} \quad  Q_3^{a_3} = 0, 2 
\ee
and so the intersection numbers are $0, \pm 1, \pm 2$.

\subsubsection*{Bulk anti-invariant three-cycles}

Let us now consider the fractional three-cycles of the form 
\be
[(1,0) \  (1,0)  \  (0,1)] \quad \stackrel{\Omega{\cal R}}{\longrightarrow} \quad [(1,0)  \ (1,0)  \ (0,-1)]
\ee
whose bulk piece is anti-invariant under the action of the orientifold. Hence the orientifold invariant combination
\be
[(0,1) \ (1,0) \ (1,0)] \, +\, [(0,-1) \ (1,0)  \ (1,0)]
\label{antibulk}
\ee
just contains linear combinations of twisted cycles. The intersection number with (\ref{rigid}) is now
\be
{1 \over 2} N_a \left[\pm Q_2^a m_a^2 \pm Q_3^a m_a^3 \right]
\ee
where the rules for the values of $Q_a^i$ can be deduced as before. Considering other three-cycles of this form we also obtain intersection numbers such as
\be
\begin{array}{l} \vspace*{.2cm}
{1 \over 2} N_a \left[\pm Q_1^a m_a^2 \pm Q_2^a m_a^3 \right]\\ 
{1 \over 2} N_a \left[\pm Q_1^a m_a^2 \pm Q_3^a m_a^3 \right]
\end{array}
\ee
In the following we will assume that these above set of three-cycles generate all the invariant homology classes of our toroidal orbifold.

\subsubsection*{The four-generation Pati-Salam example}

As an example let us consider the four-generation Pati-Salam model of table \ref{tab:aaa} and let us check whether there is any non-trivial Abelian discrete symmetry. The D6-brane stack $a_2$ corresponds to two D6-branes wrapping (\ref{exa2}), and so the $\IZ_2$ symmetry that arises from the fact that $N_{a_2} = 2$ corresponds to the center of the gauge group $SU(2)$. 

In addition we have that the stack $a_1$ corresponds to $4 \Pi^F_{a_1}$ with 
\be
\Pi^F_{a_1}\, =\, (1,0) \  (0,1)  \ (0,-1)
\label{exa1}
\ee
and so all the intersection numbers (\ref{Ktheory}) vanish identically. If we now consider the invariant combination (\ref{antibulk}) we see that $Q_2^{a_1} = Q_3^{a_1} = 1$, and so the intersection number with $\Pi^F_{a_1}$ can be made equal to one for some choice of signs. 

Finally, we consider the stack $a_3$ corresponding to $2 \Pi^F_{a_3}$ with 
\be
\Pi^F_{a_2}\, =\, (-3,2)  \  (-2,1)  \  (-4,1)
\label{exa1}
\ee
One can check that
\be
Q_1^{a_3} =1  \quad {\rm and} \quad Q_2^{a_2} = Q_3^{a_3} = 0 \quad {\rm or} \quad Q_2^{a_2} = Q_3^{a_3} = 2 
\ee
and so the intersection numbers arising from (\ref{Ktheory}) are
\be
2 \times \left(0, 1, 2 \right)
\ee
where the factor of 4 comes from $N_{a_3} = 2$ and it correspond to the $\IZ_2$ within $SU(2)$.

Hence, we conclude that in the Pati-Salam model at hand there is no non-trivial discrete gauge symmetry which is a remnant of the massive $U(1)$ symmetries, at least in the visible sector of the model. As we already know, there will be nevertheless non-trivial discrete flavor symmetries.

\newpage


\end{document}